\documentclass[aps,prd,reprint,groupedaddress]{revtex4-2}

\makeatletter
\let\@fnsymbol\@arabic
\makeatother

\usepackage{graphicx,amssymb,amsmath,amsthm,amsfonts,epsfig,epsf,array,subfigure}
\usepackage[usenames]{color}
\usepackage{xcolor}
\usepackage{hyperref}
\usepackage{booktabs}

\def\be{\begin{equation}}
\def\ee{\end{equation}}
\def\bea{\begin{eqnarray}}
\def\eea{\end{eqnarray}}
\def\bfla{\begin{flalign}}
\def\efla{\end{flalign}}
\def\nn{\nonumber}

\def\gsim{\, \rlap{$>$}{\lower 1.1ex\hbox{$\sim$}}\,}
\def\lsim{\, \rlap{$<$}{\lower 1.1ex\hbox{$\sim$}}\,}

\definecolor{purple}{rgb}{0.7,0,1}

\begin{document}

\title{The impact of higher derivative corrections to General Relativity\\on black hole mergers}

\author{Jo\~{a}o M. Dias,$^{1} $Antonia M. Frassino,$^{2,3}$ David~C.~Lopes,$^{1}$ Valentin D. Paccoia,$^{4}$ and Jorge~V.~Rocha$^{5,1,6}$}

\affiliation{
$^1$ Departamento de F\'isica, Instituto Superior T\'ecnico -- IST, \\
Universidade de Lisboa - UL,
Av. Rovisco Pais 1, 1049-001 Lisboa, Portugal\\
$^2$ Departamento de F\'{i}sica y Matem\'{a}ticas, Universidad de Alcal\'{a}, Campus universitario 28805, Alcal\'a de Henares (Madrid), Spain\\
$^3$Departament de F{\'\i}sica Qu\`antica i Astrof\'{\i}sica, Institut de Ci\`encies del Cosmos,\\ Universitat de Barcelona, Mart\'{\i} i Franqu\`es 1, E-08028 Barcelona, Spain\\
$^4$Dipartimento di Fisica e Geologia, Università degli Studi di Perugia, Via A. Pascoli, 06123, Perugia, Italy\\
$^5$Departamento de Matem\'atica, ISCTE -- Instituto Universit\'ario de Lisboa, Avenida das For\c{c}as Armadas, 1649-026 Lisboa, Portugal\\
$^6$Instituto de Telecomunica\c{c}\~oes--IUL, Avenida das For\c{c}as Armadas, 1649-026 Lisboa, Portugal}

\date{\today}

\begin{abstract}
The merging of two black holes is a notoriously difficult process to describe exactly.
Nevertheless, the hindrances posed by gravity's nonlinearity can be circumvented by focusing on the strict extreme mass ratio limit, in which one of the black holes is infinitely larger than the other. Such an approach has been developed by Emparan and Mart\'inez and applied within General Relativity to investigate the time evolution of event horizons melding, using nothing but elementary concepts in gravitational physics and simple integrations of geodesics.
We apply this strategy to study black hole mergers in higher derivative gravity, in order to assess how the defining characteristics of the fusion process change as the gravitational theory is modified. We adopt the case of Einsteinian cubic gravity for concreteness, and determine how the mergers' duration and the relative area increment change as the theory's single coupling parameter is varied. 
\end{abstract}

%\keywords{}

%\pacs{04.50.-h,04.70.-s,04.70.Bw,04.70.Dy}
%14.80.-j 	Other particles (including hypothetical)
%11.10.St 	Bound and unstable states; Bethe-Salpeter equations
%12.60.-i 	Models beyond the standard model (for unified field theories, see 12.10.-g)
%04.25.D-    Numerical relativity
%04.25.dc    Numerical studies of critical behavior, singularities, and cosmic censorship
%04.25.dg    Numerical studies of black holes and black-hole binaries
%04.25.-g    general relativity: approximation methods, equations of motion
%04.40.-b 	Self-gravitating systems; continuous media and classical fields in curved spacetime
%04.50.-h    Higher-dimensional gravity and other theories of gravity
%04.50.Cd    Kaluza-Klein theories
%04.50.Gh    Higher-dimensional black holes, black strings, and related objects
%04.60.Cf    Gravitational aspects of string theory
%04.70.-s    Physics of black holes
%04.70.Bw    Classical black holes
%04.70.Dy    Quantum aspects of black holes, evaporation, thermodynamics
%04.80.-y    Experimental studies of gravity
%04.80.Cc    Experimental tests of gravitational theories
%11.25.Mj    Compactification and four-dimensional models
%11.10.Kk    Field theories in dimensions other than four

\maketitle

%%%%%%%%%%%%%%%%%%%%%%%%%%%%%%%%%%%%%%%
\section{Introduction\label{sec:Intro}}
%%%%%%%%%%%%%%%%%%%%%%%%%%%%%%%%%%%%%%%

The remarkable scientific advances of the past seven decades, both at the theoretical and technological level, have enabled the routine detection of gravitational waves (GW)~\cite{Cervantes_Cota_2016}, beginning with the early work of Pirani~\cite{Pirani:1956tn}.
The waveforms analyzed so far validate General Relativity (GR) as an excellent approximation to the gravitational interactions we observe in nature. Yet, GR falls short of providing a complete description of quantum gravity. Therefore, modifications of GR are warranted~\cite{Clifton:2011jh, Berti:2015itd}.

A multitude of viable modifications of GR have been significantly constrained, mostly from weak field tests (e.g. Solar system and binary pulsar constraints~\cite{Will:2018bme}, or limits derived from the propagation speed of GWs~\cite{CANTATA:2021ktz}). Nevertheless, the biggest departures from GR are expected in the strong field regime, which is naturally probed in the process of compact binary mergers. 
Gravitational waveforms from such events have characteristic profiles that are typically described as the sequence of an inspiral phase, a coalescence phase, and a final ringdown phase. 
Theoretical methods have been developed to address each of these stages in order to construct the GW templates necessary to identify and properly characterize detected signals~\cite{Yunes:2009ke}. 

While the inspiral and ringdown phases of a GW signal are traditionally investigated with perturbative techniques~\footnote{For instance, the post-Newtonian approach is commonly used to study the inspiral phase, whereas the quasinormal mode analysis is well-suited to address the ringdown phase.}, which can be adapted to modified theories of gravity, the study of the highly nonlinear coalescence phase typically relies on numerical relativity or self-force methods, depending on the hierarchy of masses involved. 
For the study of mergers in modified gravity, this state of affairs is somewhat restraining, given that little is known about well-posedness and suitable formulations of the equations of motion for numerical simulations in extensions of GR~\cite{Kovacs:2020ywu}. Simultaneously, the self-force program has been developed to a great extent heavily based on GR~\cite{Barack:2018yvs}.

It is important to stress that the nonlinear coalescence phase generates the largest amplitudes of the GW signal and is therefore the best place to probe strong gravity effects. 

In this paper, we present, as proof of principle, a study of the full nonlinear merger between black holes in modified gravity employing a technique previously developed by Emparan and Mart\'inez~\cite{Emparan:2016ylg}.
The approach we will adopt allows us to circumvent the aforementioned challenges faced by numerical relativity, by relying also on a perturbative approach --- namely we shall rely on the extreme mass ratio (EMR) limit, just like in the self-force program. Therefore, this study is most relevant for future detections with LISA \cite{Colpi:2024xhw}. The main difference with respect to the self-force framework is that we will be interested in resolving scales of the order of the small object.

A compelling class of viable extensions of Einstein gravity is comprised by higher derivative gravity, where the Einstein-Hilbert action is complemented with diffeomorphism-invariant terms that involve more than two derivatives of the metric. Generically, the resulting equations of motion end up being differential equations of order higher than two, with the notable exception of Lovelock theories~\cite{Lovelock:1971yv}.
Higher curvature corrections to GR naturally appear in the context of low-energy effective theory~\cite{Donoghue:1994dn} and are also motivated by ultraviolet completions of GR ~\cite{Stelle:1976gc}. In addition, modified gravity theories have been proposed to accommodate cosmological data~\cite{DeFelice:2010aj}.
Given the long-standing interest in higher derivative gravity, there is a vast literature about black hole solutions in these theories. Aspects of  static spherically symmetric solutions of higher derivative gravity have been studied in  \cite{Holdom:2002xy, Mignemi:1991wa,Nelson:2010ig,Bueno:2017sui,Goldstein:2017rxn}. A generalization of Lovelock gravity with the addition of interactions cubic in the curvature in spacetime dimensions $D\geq5$ has been of great interest in the context of AdS/CFT \cite{Myers:2010ru,Myers:2010jv}, while solutions in Einstein gravity with quadratic curvature terms in the action have been studied in \cite{Lu:2015cqa,Lu:2015psa}.

For our case study, we will adopt Einsteinian cubic gravity (ECG), which is one such higher curvature theory~\cite{Bueno:2016xff,Hennigar:2017ego,Arciniega:2018fxj,Arciniega:2018tnn,Hennigar:2018hza,Poshteh:2018wqy}. It stands out as the most general diffeomorphism-invariant metric theory whose Lagrangian is constructed out of products of the Riemann tensor up to cubic order, with the property that its linearized spectrum on maximally symmetric backgrounds exactly matches that of GR, and whose coefficients in the Lagrangian are dimension-independent. 

Our choice of Einsteinian cubic gravity among all higher curvature gravity theories is dictated by two criteria: (i) its astrophysical viability, tied to the fact that it matches with the linearized Einstein equations around flat space, and (ii) its simplicity in several aspects. 
This simplicity is manifest in the fact that ECG only has one additional dimensionful coupling constant, $\lambda$, and that static and spherically symmetric solutions are determined by a single ordinary differential equation~\cite{Bueno:2016lrh, Hennigar:2016gkm,Frassino:2020zuv}.
Hence, ECG affords a concrete and controllable setting in which the effects of higher derivative corrections can be probed while considering simple static and spherically symmetric BHs.
Being a viable and well-motivated theory of gravity, it is important to test some of its predictions against observations; for instance, such confrontation has allowed to put upper bounds on the coupling $\lambda$ by analyzing the deviation it would cause on BH shadows~\cite{Hennigar:2018hza}.
We could have equally considered quadratic gravity but: (i) its static and spherically symmetric solutions are determined by a coupled systems of ODEs; (ii) apart from a (thermodynamically unstable) branch of solutions, the static spherically symmetric BHs are exactly the same as in GR, i.e. the Schwarzschild spacetime; (iii) purely quadratic terms in the Lagrangian can be eliminated by a field redefinition~\cite{deRham:2020ejn} (at the expense of introducing non-minimal couplings with matter fields and higher dimension operators in the spirit of effective field theories).

\medskip

The goal of this paper is to quantify the influence of higher curvature corrections on extreme mass ratio BH mergers. Namely, we aim to determine how the duration of the process and the increase in the relative area vary as functions of the ECG coupling constant. The duration of a binary merger is a quantity of primordial relevance for gravitational waves, so it is expected our results can be translated into constraints on $\lambda$ from gravitational wave detections.

To characterize the extreme mass ratio binary merger we will resort to the ray-tracing techniques first developed in this context in Ref.~\cite{Emparan:2016ylg} to study the fusion of two BHs, the smaller one being of Schwarzschild form, and later employed in~\cite{Emparan:2017vyp, Emparan:2020uvt, Pina:2022dye, Dias:2023pdx} to analyze a variety of situations.
To apply this program in a modified gravity setup (respecting the equivalence principle) we only need to have closed form solutions for the spacetime describing a static black hole in the theory considered. Exact analytical solutions in ECG are out of reach, so we will resort to an analytic approximation scheme.

The method of~\cite{Emparan:2016ylg} yields quantitative information about some of the effects of modifications of GR on the binary merger, namely on the evolution of the event horizon.
However, we should remark that at this stage the effects on the GW signal are assessed only qualitatively. The technique used requires further development (in progress) in order to make quantitative predictions about GW signals.

The paper is organized as follows: In Section~\ref{sec:ECG} we describe black holes in ECG and the determination of such solutions by using the continued fraction method. In Section~\ref{sec:Mergers} we apply the Emparan-Mart\'inez technique to study BH mergers with the previously determined background. Finally, in Section~\ref{sec:Conclusion} we draw our conclusions and discussion.

%%%%%%%%%%%%
\section{Black holes in Einsteinian cubic gravity\label{sec:ECG}}
%%%%%%%%%%%%

Four-dimensional ECG is defined by the following Lagrangian density, 
\be
{\cal L} = \frac{1}{16\pi G} \left(R - \lambda G^2{\cal P} + {\cal O}\right)\,,
\label{eq:Lagrangian}
\ee
where $G$ is the Newton gravitational constant and $\lambda$ is the additional coupling constant of the theory. It generalizes the usual Einstein-Hilbert Lagrangian, $16\pi G{\cal L} = R$, where $R$ stands for the Ricci scalar, by adding a certain combination of cubic curvature terms, defined by
\bea
{\cal P} &\equiv& 12 {{{R_a}^c}_b}^d {{{R_c}^e}_d}^f {{{R_e}^a}_f}^b
+ {R_{ab}}^{cd} {R_{cd}}^{ef} {R_{ef}}^{ab} \nn\\
&& -12 R_{abcd} R^{ac} R^{bd}
+8 {R_a}^b {R_b}^c {R_c}^a\,.
\eea
The remaining invariant ${\cal O}$ in Eq.~\eqref{eq:Lagrangian}
gathers terms that are topological in nature, as well as terms that do not contribute to the equations of motion when restricting to a spherically symmetric line element, as we will do. For our purposes, we can simply drop this term from the Lagrangian.

We will assume $\lambda\geq0$ throughout. This ensures that the theory has a unique vacuum~\cite{Bueno:2016xff} and that the static and spherically symmetric BHs we shall study are regular on and outside the horizon~\cite{Frassino:2020zuv}.

The generalized Einstein equations are derived from the Lagrangian~\eqref{eq:Lagrangian}, yielding
\be
{\cal E}_{ab} \equiv E_{acde}{R_b}^{cde} - \frac{1}{2}g_{ab}{\cal L} - 2\nabla^c\nabla^d E_{acdb} = 0 \,,
\label{eq:EFE}
\ee
where $E^{abcd} \equiv \partial {\cal L} / \partial R_{abcd}$.

Static and spherically symmetric black hole solutions of ECG are characterized by a line element  of general form~\cite{Hennigar:2016gkm,Bueno:2016lrh}
\be
ds^2 = - f(r) dt^2 + f(r)^{-1}dr^2 + r^2 d\Omega^2\,.
\label{eq:metric}
\ee
This is the exact same form as that of Schwarzschild. However, in ECG the blackening factor $f(r)$ is not simply a polynomial in inverse powers of the radial coordinate, $r$. Instead, the field equations~\eqref{eq:EFE} demand that it must satisfy a nonlinear ordinary differential equation,
%
%%%%%%%%
\begin{widetext}
\be
-(f-1)r - G^2\lambda \left[ 4f'^3 +12\frac{f'^2}{r} - 24f(f-1) \frac{f'}{r^2} - 12f f''\left(f'-\frac{2(f-1)}{r}\right) \right] = 2GM \,.
\label{eq:EOM}
\ee
\end{widetext}
%%%%%%%%
%
A spacetime of the form~\eqref{eq:metric} with a metric function $f$ obeying this sole equation of motion automatically solves all the components of the generalized Einstein equations~\eqref{eq:EFE} ---of which equation~\eqref{eq:EOM} is in fact a first integral.
From this point of view, the parameter $M$ is just a constant of integration, but physically it  corresponds to the gravitational mass of the spacetime.

It is easy to see that the Schwarzschild solution is recovered when the coupling constant is set to $\lambda=0$. However, for $\lambda\neq0$ closed form solutions of equation~\eqref{eq:EOM} are unfortunately out of reach. Therefore, one must either resort to numerical integration or to  approximation schemes.

From now on we will work with natural units, for which $G=1$. 

Let us first briefly describe the typical numerical integration approach.
An asymptotic analysis of~\eqref{eq:EOM} shows that two kinds of corrections to the Schwarzschild solution $f_0(r)= 1-2M/r$ are turned on when $\lambda\neq0$~\cite{Bueno:2016lrh}. In addition to perturbative corrections in $\lambda$, which appear multiplying (high order) inverse powers of $r$, there are two non-perturbative corrections, schematically of the form $\sim \exp\left(\pm\frac{r^{5/2}}{\sqrt{\lambda}}\right)$. Upon demanding the absence of the growing exponential term, we are left with a general asymptotically flat solution that is controlled by a single free parameter, namely the coefficient of the decaying exponential.

An equivalent analysis can be performed at the other `end' of the black hole exterior: the horizon radius, $r_h$, which is the outermost root of the blackening factor, $f(r_h)=0$. This value of $r$ happens to be a singular point of the differential equation~\eqref{eq:EOM}. Nevertheless, by assuming regularity of $f$ and performing a series expansion around such a point, one can solve~\eqref{eq:EOM} order by order in powers of $(r-r_h)$. It turns out that {\it all} the coefficients of this series can be determined as a function of the second Taylor coefficient. Therefore, the behavior of the solution around the horizon is also controlled by a single free parameter.

A global black hole solution that is regular on and outside the horizon can then be obtained by `shooting' from the two extremes and imposing continuity and differentiability of the function at some intermediate radius. These two conditions completely fix the two free parameters of the local solutions near the horizon and in the asymptotic region, leaving a unique black hole solution for each value of $\lambda$. In practice, what we do is to start the integration from the horizon and vary the free parameter controlling the behavior around $r_h$ so as to maximize the value of $r$ at which the numerical solution diverges due to the growing exponential mentioned above. Ideally, this would happen at $r=+\infty$.

While the above approach is well-defined and can be employed, in principle, to obtain asymptotically flat black holes, in practice the results have inherent shortcomings: the validity of the solution thus obtained is limited to a relatively short radial interval. (More than $10$ times the horizon radius is already challenging to achieve.) This is due to our inability to fix the free parameter at the horizon with infinite precision, as well as the unavoidable numerical inaccuracies in the integration scheme. Both of these effects inevitably activate the growing exponential term present in the general asymptotic solution, rendering the numerical solution singular ---and definitely not asymptotically flat.

An alternative approach, which better suits our purposes, is the use of the continued fraction method to approximate a true solution of~\eqref{eq:EOM}. This idea was first exploited in the present context in Ref.~\cite{Hennigar:2018hza} and has the advantage of producing good approximations that are valid on the entire range $r\geq r_h$. This is extremely important for our intents, since we want to integrate expressions that depend on the blackening factor $f(r)$ over a range that extends far beyond the horizon radius of the small black hole.
Thus, we now review how the continued fraction method is applied within this setting.

To begin with, it is convenient to define a new coordinate $x=1-r_h/r$, which maps the exterior of the black hole, $r \in [r_h,\infty)$, to the finite interval $x\in[0,1)$. A suitable ansatz for $f$ that smoothly interpolates between zero at the horizon $r=r_h$ and unity at $r\to \infty$ can be written as
\be
f(x) = x \!\left[ 1 \!- \!\varepsilon(1\!-\! x) \!+ \!(b_0\!-\!\varepsilon)(1\!-\!x)^2 \!+\! B(x)(1\!-\! x)^3 \right],
\label{eq:contfrac1}
\ee
where $B(x)$ is expressed as a continued fraction,
\be
B(x) = \frac{b_1}{1+\frac{b_2 x}{1+\frac{b_3 x}{1+\dots}}}\,.
\label{eq:contfrac2}
\ee

The asymptotic behavior at $x=1$ fixes the coefficients $\varepsilon$, $b_0$ and $b_1$ to be
\bea
\varepsilon &=& \frac{2M}{r_h}-1\,,\\
b_0 &=& 0\,,\\
b_1 &=& \frac{2r_h^2 \kappa_g - 3 r_h + 4M}{r_h}\,.
\eea
where $\kappa_g = f'(r_h)/2$ represents the black hole's surface gravity.

While the coefficient $b_2$ cannot be determined from a near-horizon analysis, all the higher order coefficients $b_n$, with $n>2$, get fixed once $\lambda/r_h^4$ and $b_2$ are selected. Indeed, note that the dimensionless quantities $M/r_h$ and $\kappa_g r_h$ are uniquely determined for each choice of $\lambda/r_h^4$ by~\cite{Bueno:2016lrh}
\bea
\kappa_g r_h &=&  \frac{1}{1+\sqrt{1+48 \lambda /r_h^4}}\,,\label{eq:kg}\\
\frac{2M}{r_h} &=& 1-\frac{16\lambda}{r_h^4}\frac{5+3\sqrt{1+48 \lambda /r_h^4}}{\left(1+\sqrt{1+48 \lambda /r_h^4}\right)^3}\,\label{eq:mass}.
\eea
The coefficient $b_2$ can be determined in two different ways, as discussed in \cite{Hennigar:2018hza}. One approach is to truncate the continued fraction~\eqref{eq:contfrac2} at some order $n$, so that $b_{n+1}(b_2)=0$, which can be solved for $b_2$. 
This has some drawbacks. The value of $b_2$ depends on the truncation order and it's not clear that it converges at higher orders. Furthermore, the solution to equation $b_{n+1}(b_2)=0$ is typically not unique.

Another approach is to numerically solve the differential equation~\eqref{eq:EOM}, starting from a point near the horizon, $r=r_h(1+\epsilon)$, with $\epsilon \ll 1$, where we impose the following initial conditions:
\bea    
f(r_h(1+\epsilon))=2 \kappa_g r_h\epsilon + \sum_{j=2}^{n} a_j (r_h \epsilon)^j \label{eq:TAYLOR1} \,,\\
f'(r_h(1+\epsilon))=2 \kappa_g + \sum_{j=2}^{n} j a_j (r_h \epsilon)^{j-1}\,.
\label{eq:TAYLOR2}
\eea
By comparing Eq.~\eqref{eq:TAYLOR1} with a similar series expansion of Eq.~\eqref{eq:contfrac1} around the horizon, we obtain the coefficient $b_2$ uniquely define by the parameter $a_2$:
\be
b_2=\frac{6r_h -6M-8 \kappa_g  r_h^2  - r_h^3 a_2}{4M-3r_h+ 2 \kappa_g r_h^2 }\,,
\label{eq:b2_vs_a2}
\ee
whereas the $a_j$'s with $j>2$ are recursively obtained as a function of $a_2$.

The value of $a_2$ that is needed to plug in Eq.~\eqref{eq:b2_vs_a2} is uniquely determined by requiring that the numerical solution for $f(r)$ is asymptotically flat. In practice, this can never be achieved, as the finite precision of the integration scheme inevitably triggers the exponentially growing mode at a given radius $r_s(a_2)$ which depends on the choice of $a_2$. Thus, the desired value of $a_2$ is $a_2= \arg \max_{\alpha} r_s(\alpha)$, which maximizes the domain in which the numerical solution is valid.

%
%%%%%%%%%
\begin{figure}
\includegraphics[width=8.3cm]{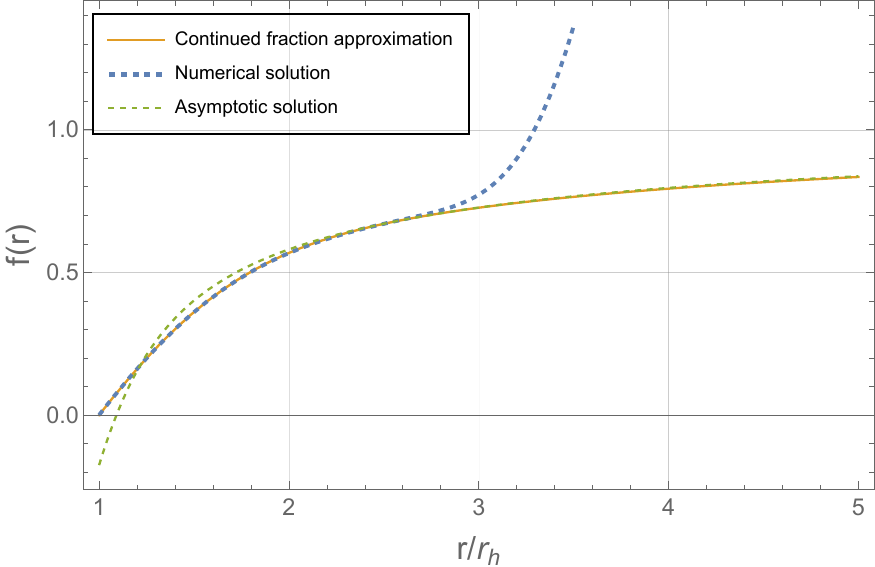}
\caption{Solution for $f(r)$ obtained in three different ways: using the ninth-order ($b_{10}=0$) continued fraction expansion~\eqref{eq:contfrac1} (orange and thick line); numerically integrating the differential equation~\eqref{eq:EOM} (blue and dashed line); using the asymptotic expression~\cite{Bueno:2016lrh} $f(r)=1-2M/r-432 \lambda  M^2/r^6 -736 \lambda  M^3/r^7$ (green and dashed line). The results presented are for $\lambda/r_h^4=0.0164738$, which in mass units corresponds to $\lambda/M^4=0.6$\,. 
\label{fig:num_vs_cf_vs_asympt}}
\end{figure}
%%%%%%%%%
%

As an example, in Fig.~\ref{fig:num_vs_cf_vs_asympt} we present the solution for the blackening factor $f$ obtained by numerically integrating the differential equation~\eqref{eq:EOM} for a specific value of $\lambda/r_h^4$, together with its asymptotic expansion and the continued fraction approximation.
By construction, the numerical solution vanishes at the horizon, and for somewhat larger radii it shows regular and monotonic behavior. However, at a certain radius $r_s$ the numerical solution blows up, signaling the dominance of the exponentially growing mode of the asymptotic solution \footnote{Note that $r_s/r_h$ tends to be larger for larger values of $\lambda/r_h^4$, and so the numerical solution is valid in a wider range as we increase the coupling parameter. This is because, as we increase $\lambda/r_h^4$, we decrease the exponent of the asymptotic behavior $\sim \exp \left(\frac{(r/r_h)^{5/2}}{\lambda/r_h^4}\right)$, which delays its contribution to the numerical solution.}.
As can be seen from the continuous line in Fig.~\ref{fig:num_vs_cf_vs_asympt}, the continued fraction expansion~\eqref{eq:contfrac1}  smoothly transitions between the near-horizon solution and the asymptotic solution, appropriately describing $f(r)$ in the whole radial domain.
%
%%%%%%%%%
\begin{figure}
\includegraphics[width=8.3cm]{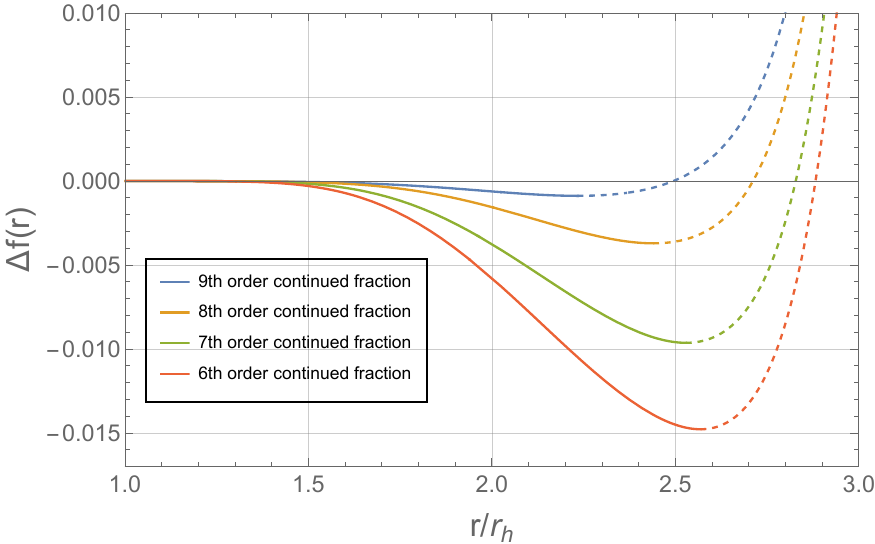}
\caption{Difference between the numerical solution and the continued fraction solution, truncated at ninth-order (blue), eighth-order (orange), seventh-order (green) and sixth-order (red), i.e. the truncation order decreases from top to bottom. The dashed lines represent the region where $\Delta f$ starts to increase, presumably due to the inaccuracy of the numerical solution. The results presented are for $\lambda/r_h^4=0.0164738$.
\label{fig:num_vs_cf}}
\end{figure}
%%%%%%%%%
%

Fig.~\ref{fig:num_vs_cf} shows $\Delta f$, the difference between the numerical solution and the continued fraction approximation, as a function of the radius, for different truncation orders of the continued fraction expansion. For sufficiently large radii, $\Delta f$ becomes large, but this effect is entirely due to the (unphysical) divergence of the numerical solution. Therefore, we are specifically interested in the behavior of $\Delta f$ only for radii smaller than the value of $r/r_h$ where this function attains its minimum (in Fig.~\ref{fig:num_vs_cf}, this is indicated with solid lines). We observe that, within this range of radii, $|\Delta f|$ decreases as we increase the order of the expansion. In particular, the ninth-order continued fraction yields $|\min\{\Delta f\}| \sim 10^{-3}$. 

\begin{figure}
\includegraphics[width=8.5cm]{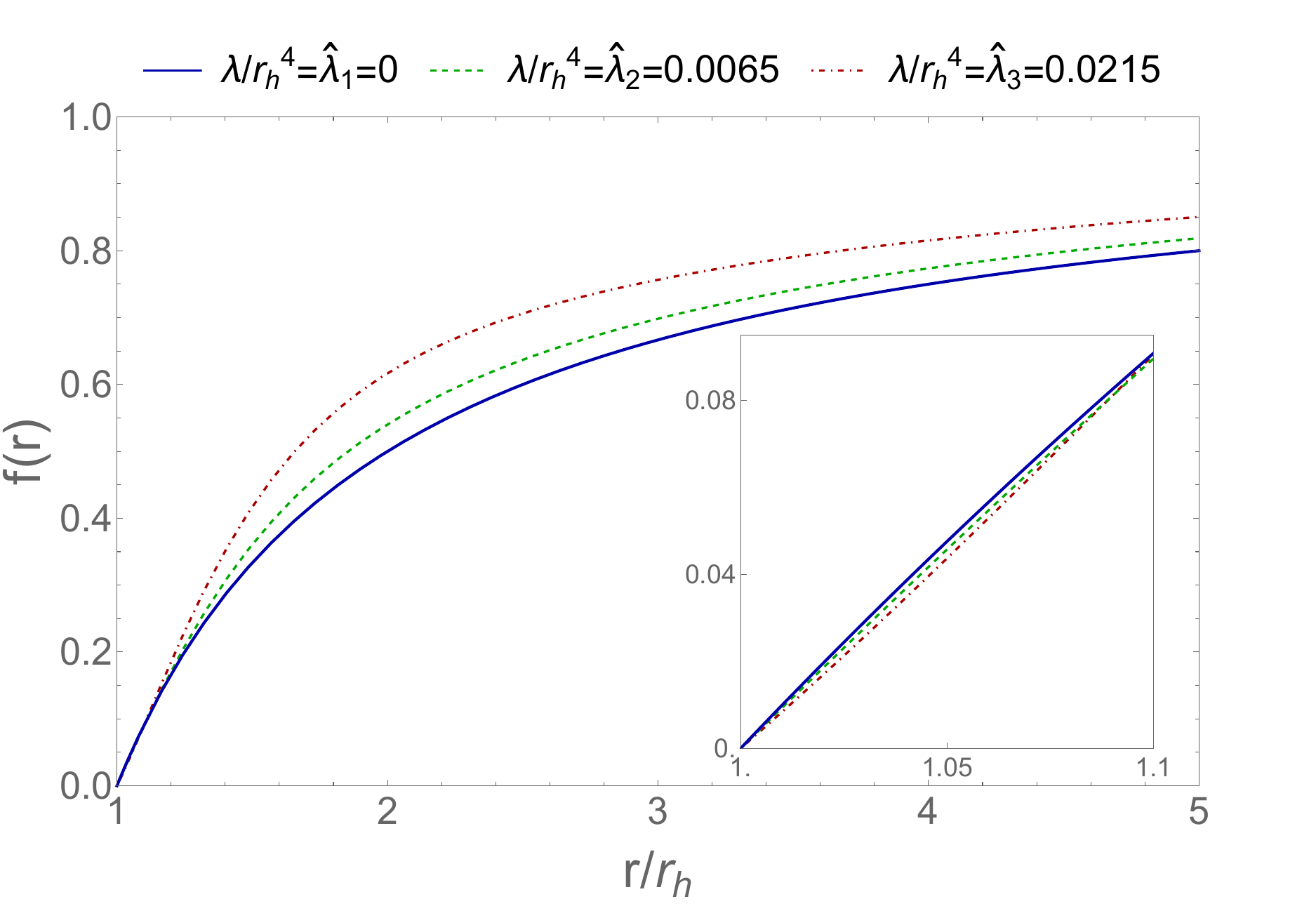}
\caption{Blackening factor $f(r)$ for $\lambda/r_h^4=0$ (blue and thick curve), i.e. the Schwarzschild case, $\lambda/r_h^4=0.0065$ (green and dashed curve) and $\lambda/r_h^4=0.0215$ (red and dashed-dotted curve). The latter two curves were obtained using a ninth-order continued fraction approximation. While smaller $\lambda$'s favor stronger gravitational fields in the near-horizon region (higher surface gravity) and in the asymptotic region (higher black hole mass), the opposite trend emerges in an intermediate region. The interplay between these two effects throughout the generators' path is what determines the characteristics of the merger for a given $\lambda$.
\label{fig:grav_pot}}
\end{figure}

The blackening factor displays the same qualitative behavior for all values of the coupling constant $\lambda$, with small quantitative changes that will impact the null geodesics that we compute in Sec.~\ref{sec:Mergers}. This is illustrated in Fig.~\ref{fig:grav_pot}, where we plot the profile $f(r)$ for three different values of $\lambda$. 
The behavior of the function $f(r)$ with $\lambda$ is very non-trivial. The surface gravity~\eqref{eq:kg} (being the derivative of $f(r)$) decreases with increasing $\lambda$, so that the gravitational field, in the near-horizon region, is weaker for larger $\lambda$. Far from the black hole horizon, the gravitational field is well approximated by the Newtonian potential. Given that the mass decreases with increasing values of $\lambda$, this implies that the gravitational field in the asymptotic region becomes weaker for larger values of $\lambda$. By continuity there is an intermediate range of radii in which the gravitational field becomes stronger with increasing $\lambda$.   
%
%%%%%%%%%
%%%%%%%%%
%

A final comment is in order.
The appearance of the coupling constant $\lambda$ introduces a further mass scale, $\lambda^{-1/4} G^{-1/2}$, in addition to the physical mass $M$ of a given black hole solution. One consequence is that BHs in ECG do not in general satisfy $r_h=2GM$: the relation between the horizon radius and the mass is affected by the coupling constant. When performing calculations, and displaying our results, it is therefore distinct to fix $M=1$ or $r_h=1$. Although one may always convert between the two normalizations, expressing results in units of the mass has the advantage that it is coordinate independent. Nevertheless, to compare the behavior of geodesics we find it much more convenient to fix $r_h=1$ for all the BHs we consider (with different couplings $\lambda$). We have therefore normalized our black hole background to have $r_h=1$, irrespective of $\lambda$, as done in Figs.~\ref{fig:num_vs_cf_vs_asympt} and~\ref{fig:num_vs_cf}.

%%%%%%%%%%%%
\section{Black hole mergers in ECG\label{sec:Mergers}}
%%%%%%%%%%%%

%%%%%%%%%%%%
\subsection{Generators of the merging horizon}
%%%%%%%%%%%%

We are interested in determining a certain class of null geodesics in the background of an ECG black hole, which will form the event horizon generators. At late times we want this congruence of geodesics to approach a null plane, each individual light ray being parallel to all others.

Let $x^\mu=(t,r,\theta,\phi)$ be the spacetime coordinates of one such geodesic, and denote by $\kappa$ the affine parameter used to distinguish points along the geodesics. The geodesic equation reads
\be
\frac{d^2 x^\mu}{d\kappa^2} + \Gamma^\mu_{\alpha\beta} \frac{d x^\alpha}{d\kappa} \frac{d x^\beta}{d\kappa} = 0 \,,
\label{eq:geodesic}
\ee
where $\Gamma^\mu_{\alpha\beta}$ stands for the Christoffel symbols.

Since the background is spherically symmetric, we may orient the coordinate system in such a way that our geodesic remains always on the `equatorial' plane. I.e., we may choose, without loss of generality, to fix the polar coordinate to $\theta(\kappa)=\pi/2$, the geodesic equation guaranteeing that the geodesic experiences no polar acceleration, $d^2\theta/d\kappa^2=0$.

The remaining coordinates of the geodesic are determined by the other components of Eq.~\eqref{eq:geodesic}, which yield second order differential equations. However, there is a shortcut which immediately outputs first order equations.
The fact that $\partial/\partial t$ and $\partial/\partial \phi$ are Killing vectors of the background spacetime allows one to directly derive expressions for two components of the 4-velocity vector:
\be
\frac{d t}{d\kappa} = \frac{E}{f(r)} \,, \qquad 
\frac{d \phi}{d\kappa} = -\frac{L}{r^2} \,.
\ee
Here, $E$ and $L$ represent the energy and the angular momentum associated with these geodesics. Imposing then the light-like condition for the null rays, $g_{\mu\nu} \dot{x}^\mu\dot{x}^\nu=0$, immediately yields the remaining component of the 4-velocity vector,
\be
\frac{d r}{d\kappa} = \pm \sqrt{E^2 - L^2 \frac{f(r)}{r^2}} \,.
\ee

It is advantageous to use the radial coordinate $r$ to describe the null generators of the horizon. To that end, we express the azimuthal angle $\phi$ and the time coordinate $t$ both as functions of $r$:
\be
\frac{d \phi}{d r} = \mp \frac{1}{r^2 \sqrt{\frac{1}{q^2} - \frac{f(r)}{r^2}}}  \,, \qquad 
\frac{d t}{d r} = \pm \frac{1}{q f(r) \sqrt{\frac{1}{q^2} - \frac{f(r)}{r^2}}} \,,
\label{eq:PHIdotTdot}
\ee
where $q:= L/E$ is the impact parameter, which will be used to distinguish between the different null generators.

\begin{figure}[t]
\includegraphics[width=8cm]{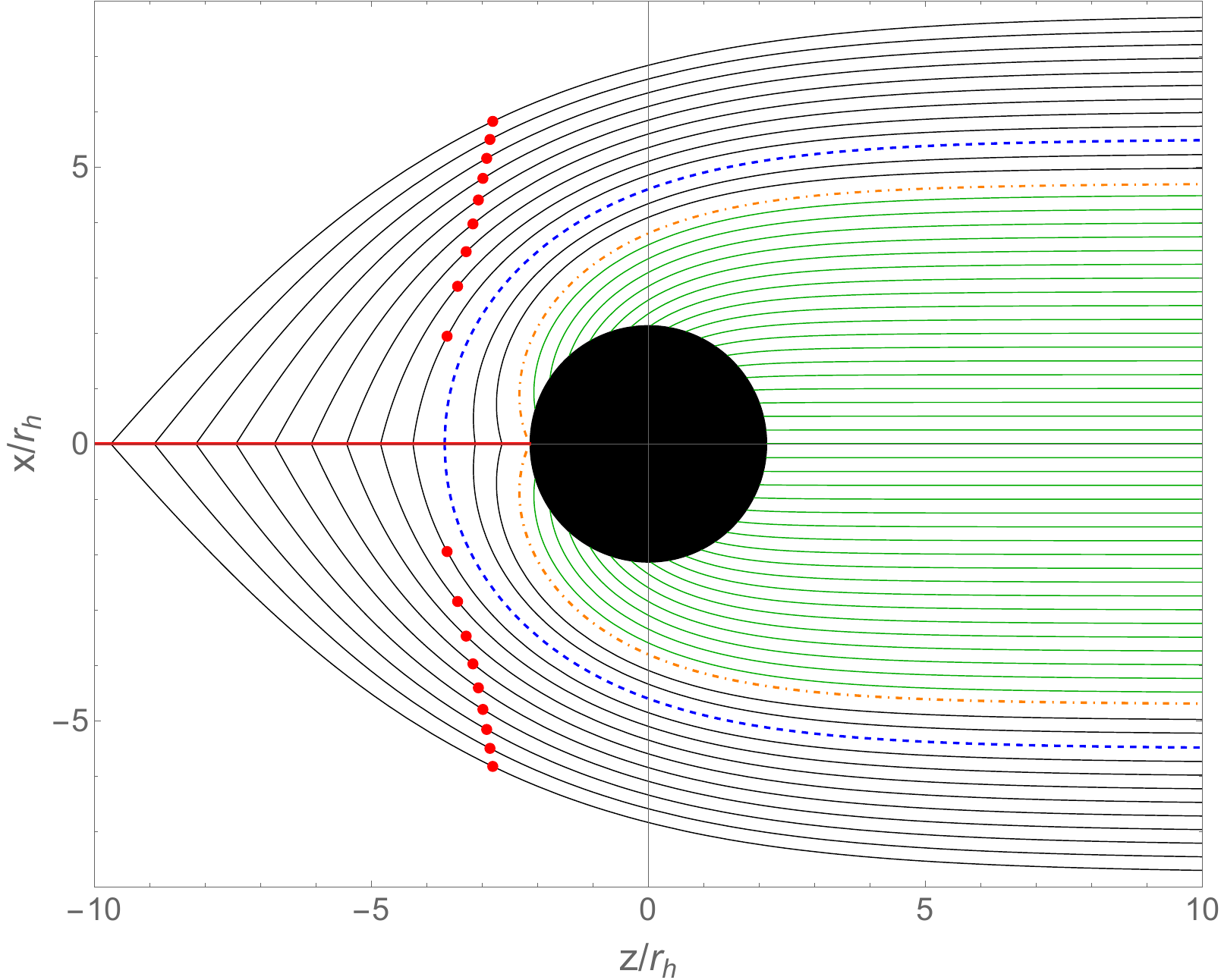}
\caption{Projection of the event horizon in the $(x,z)$-plane. Non-caustic generators with $q < q_c$ (green curves) and $q=q_c$ (orange and dashed-dotted curve) extend to $r=1$ at infinitely early times. Caustic generators (black curves) have $q>q_c$ and hit the caustic line at $\phi = \pi$ (red line). The generator with $q=q_*$ (blue and dashed curve) separates the caustic generators that bend towards the small black hole ($q_c<q<q_*$), from the caustic generators that bend away from it ($q>q_*$). In the latter case, the generators reach turning points (red dots), which are their closest points to the small black hole. The results presented are for $\lambda/r_h^4=0.00473034$, which in mass units corresponds to $\lambda/M^4=0.1$.
\label{fig:xvsz}}
\end{figure}

The equations above, in this form, make it manifest that the geodesics have a turning point (a point at which $r$ remains momentarily constant along the geodesic, while $\phi$ and $t$ continue varying) at an $r$-coordinate, $r_t$, such that
\be
\frac{f(r_t)}{r_t^2} = \frac{1}{q^2}\,.
\ee
Moreover, given that the blackening factor $f$ is a monotonically increasing function that asymptotes to $1$, it is clear that geodesics with sufficiently small impact parameter will not possess any turning point: these are the generators that connect (in the far past) with the horizon of the small black hole, or those that are born on the `small BH side' of the caustic line (see later description of Fig.~\ref{fig:xvsz}).

%
%%%%%%%%%
%%%%%%%%%
%

In order to determine the event horizon of the extreme mass ratio merger, we must then compute the full development of a null geodesic congruence that asymptotes a null plane at late times (being `null' implies that the radial coordinate must diverge when $t\to\infty$). This is done by integrating Eqs.~\eqref{eq:PHIdotTdot} backwards in $r$ for a variety of null geodesics, which are labelled by their impact parameter $q$. The initial condition for each of these integrations is set by the requirement that the integral curves all approach the same null plane, which amounts to demanding that~\cite{Emparan:2016ylg}
\bea
\phi_q(r\to\infty) &=& \frac{q}{r} + O(r^{-3})\,, \\
t_q(r\to\infty) &=& r + 2M \ln(r/2M) + O(r^{-1})\,.
\eea
The numerical integration is performed until the null geodesic either:
\begin{itemize}
\item[(a)] reaches the small black hole horizon, $r=1$, which can only happen when $t\to -\infty$. These geodesics will be referred to as the non-caustic generators of the event horizon.
\item[(b)] hits the caustic line at $\phi=\pi$, at which moment we immediately stop the integration. These geodesics constitute the caustic generators of the event horizon.
\item[(c)] goes off to a large radius after having gone through a point of closest approach to the small black hole. This can happen only if the geodesic bends {\it away} from the central object. This phenomenon does not occur with the kind of backgrounds we are considering, but it is a feature that appears in more general situations, for example when the background spacetime is taken to be an over-extreme Reissner-Nordstr\"om naked singularity.
\end{itemize}
The caustic line physically represents a collection of points in the spacetime from where more than one light ray can be emitted ---in different directions--- to reach the same asymptotic null plane. More details about the non-smooth features of the event horizon evolution during the merging of the two black holes in the EMR can be found in Refs.~\cite{Gadioux:2023pmw,Gadioux:2024tlm}.

The distinction between the caustic and non-caustic generators is illustrated in Fig.~\ref{fig:xvsz}, which shows a projection of the event horizon generators on the $(x,z)$ spatial plane, with $x=r\sin \phi$ and $z=r\cos\phi$, for a specific value of the coupling constant. As can be seen, there is a critical generator with impact parameter $q=q_c$ that separates the non-caustic generators, with $q \leq q_c$, from the caustic generators, with $q>q_c$.
The latter geodesics can be further differentiated: generators with $q_c < q < q_*$ approach the caustic line from the side of the small black hole (i.e. they bend towards the small black hole), while generators with $q>q_*$ have a turning point, approaching the caustic line from the side of the large black hole (i.e. they escape the small black hole's pull).

Since a turning point is a singular point of Eqs.~\eqref{eq:PHIdotTdot}, the numerical integration halts for generators with $q\!>\!q_*$, but that does not mean the geodesic is incomplete. It is just a consequence of using the radial coordinate as the integration variable instead of the affine parameter. There is, however, a simple way to circumvent this problem. Because the background is spherically symmetric, such a geodesic must be symmetric around a turning point. Therefore one may just perform ``half'' of the integration and then glue the other symmetric half by imposing continuity of the geodesic functions and their derivatives~\cite{Emparan:2016ylg}.

%
%%%%%%%%%
\begin{figure}[t]
\includegraphics[width=9cm]{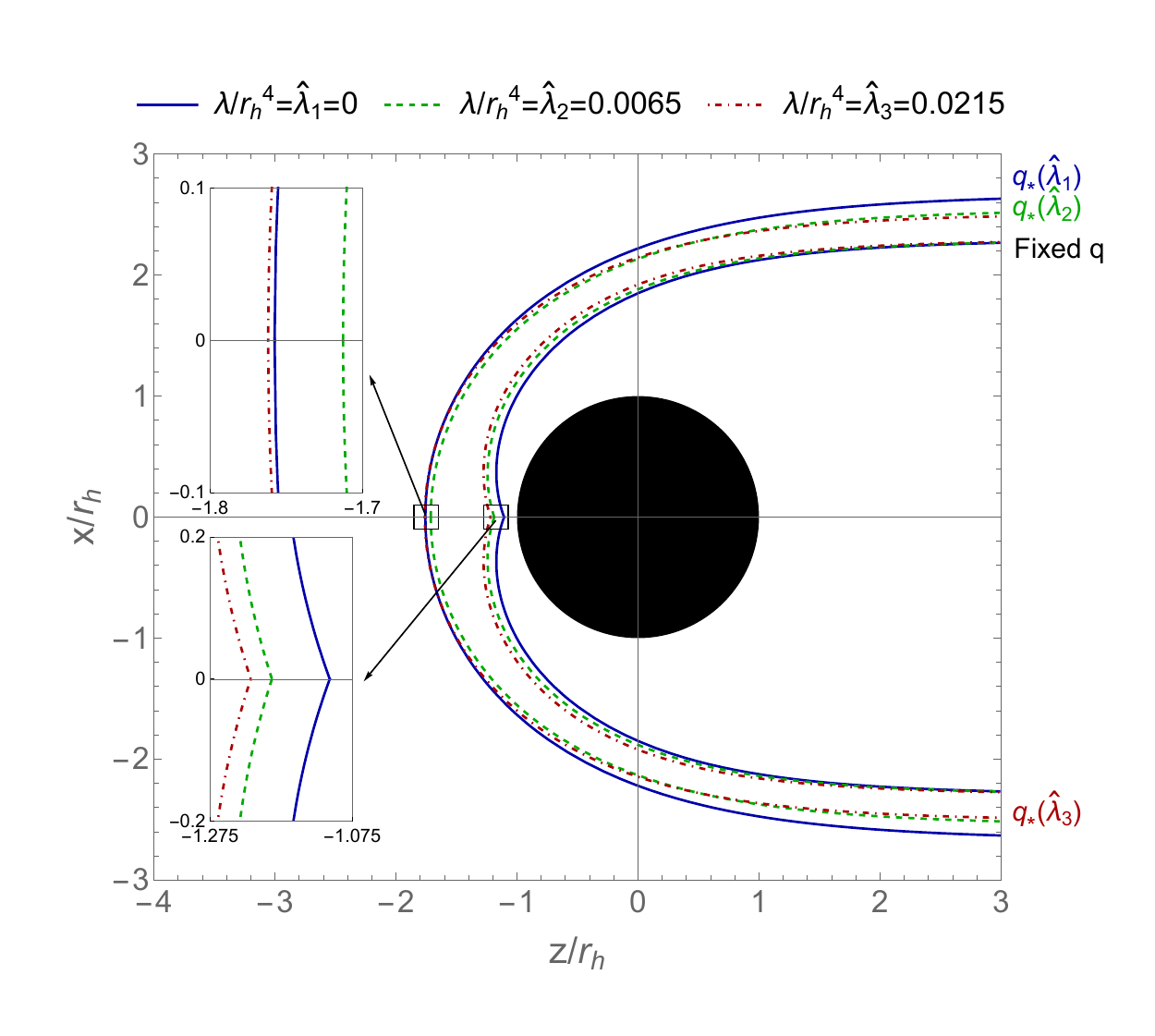}
\caption{Projection of generators of the event horizon in the $(x,z)$-plane, for $\lambda/r_h^4=0$ (blue and thick curves), $\lambda/r_h^4=0.0065$ (green and dashed curves) and $\lambda/r_h^4=0.0215$ (red and dashed-dotted curves), the same choices made for Fig.~\ref{fig:grav_pot}. The three innermost generators share the same impact parameter $q$. A zoomed-in view of these generators near the caustic is shown in the bottom left subfigure --- smaller coupling parameters favor the deflection of generators with the same impact parameter. The remaining three generators have impact parameter $q=q_*$. A zoomed-in view of these generators near the caustic is shown in the top left subfigure --- the radial distortion of the small black hole, as well as $r_*$, varies non-monotonically with $\lambda$.
Note that all the generators with $|q|<q_*$ asymptote a disk at infinity. All the generators that either connect to the small BH horizon at the infinite past or that enter the event horizon at the caustic on the side of the small BH, end on this disk.}
\label{fig:xvsz2}
\end{figure}
%%%%%%%%%
%

Figure~\ref{fig:xvsz2} shows some horizon generators for the same values of $\lambda$ adopted in Fig.~\ref{fig:grav_pot}. This is done by keeping fixed $r_h=1$.
As $\lambda$ is increased, geodesics with the same impact parameter $q$ are less deflected towards the small black hole. Naively, this should come as no surprise, given that both the mass and the surface gravity of the black hole decrease with the growth of the coupling parameter (see Eqs.~\eqref{eq:kg} and~\eqref{eq:mass}). 

Consider now the generator with $q=q_*$, as we vary $\lambda$. This generator originates from the crest of the caustic, at $(t,r,\phi)=(t_*,r_*,\pi)$ \cite{Emparan:2016ylg}. The value of $r_*$ gives a measure of the maximum distortion of the small black hole horizon during the merger. 
Note that the impact parameter $q_*$ decreases with $\lambda$, but
$r_*$ does not vary monotonically. In particular, there exists a specific value of $\lambda$ for which the distortion of the small black hole is minimized.
The reason for such non-monotonicity stems from the non-trivial behavior of the blackening factor along the generators' trajectory, as a function of the coupling. Each of these null geodesics crosses a very wide range of radii along their path, $r\in[r_*,+\infty)$, and the blackening factor $f(r)$ in different regions (asymptotic, intermediate, and near-horizon) shows different trends with $\lambda$, as exemplified in Fig.~\ref{fig:grav_pot}.
Technically, $r_*$ is determined by solving a coupled system of equations,
\be
f(r_*)\, q_*^2 = r_*^2\,,
\qquad
\phi(r_*)=\pi\,,
\ee
but the latter is actually an integral equation, taking into account Eq.~\eqref{eq:PHIdotTdot}. Therefore, the whole worldline of the geodesic is relevant for the calculation: the problem is non-local. 
Even knowing how the blackening factor $f(r)$ changes with the coupling ---as $\lambda$ is increased, $f$ becomes flatter for sufficiently large $r/r_h$ but steeper sufficiently close to the horizon---, it is difficult to anticipate the qualitative dependence of $r_*$ on $\lambda$. This illustrates how challenging it is to predict these merger features, and it highlights the importance of developing these methods to determine the evolution of the horizon in black hole mergers beyond General Relativity.

%%%%%%%%%%%%
\subsection{Extracting merger properties}
%%%%%%%%%%%%

The computations described in the previous subsection allow us to conduct a quantitative analysis of the effects of modifying General Relativity by the inclusion of higher derivative terms in the Lagrangian --- in the present case, these terms are cubic in the Riemann tensor. 
Specifically, for each value of the coupling constant $\lambda$ the merging horizons can be determined. From that, we extract defining characteristics of the merger, such as the duration of the process and the relative increase in the area of the event horizon. Finally, we assess how each of these properties varies with $\lambda$. 

\begin{figure}
\includegraphics[width=8cm]{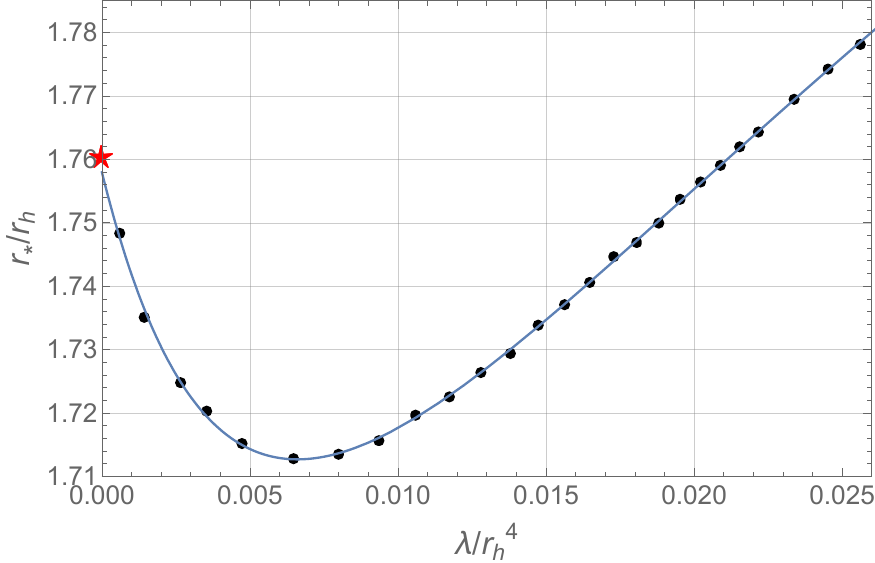}
\caption{Maximum radial distortion of the small black hole, $r_*/r_h$, as a function of the coupling constant, $\lambda/r_h^4$. The data points were fitted to a Padé approximant with six free parameters. The red star, $r_*=1.76031\, r_h$, corresponds to the value obtained by Emparan and Mart\'inez in \cite{Emparan:2016ylg} for a Schwarzschild black hole ($\lambda/r_h^4=0$).
\label{fig:rstar}}
\end{figure}

Recall that the maximum distortion of the event horizon is determined by the generator with $q=q_*$, which reaches the caustic at $r=r_*$ and $t=t_*$, the pinch-on instant. 
Figure~\ref{fig:rstar} shows how $r_*$ varies non-monotonically as a function of the theory's coupling constant, $\lambda$. In the range of the coupling parameters analyzed, $\lambda/r_h^4 = 0.0066$ yields the value for which the distortion is minimized. At its minimum, $r_*=1.71270\, r_h$. As argued in the last subsection, this non-monotonicity stems from the non-trivial behavior of the blackening factor along the generators' trajectory, as a function of the coupling. We note, however, that if we had fixed the mass $M$ instead of $r_h$, $r_*$ would be monotonically increasing as a function of $\lambda/M^4$.
%
%%%%%%%%%
%%%%%%%%%
%

\begin{figure}
\includegraphics[width=8cm]{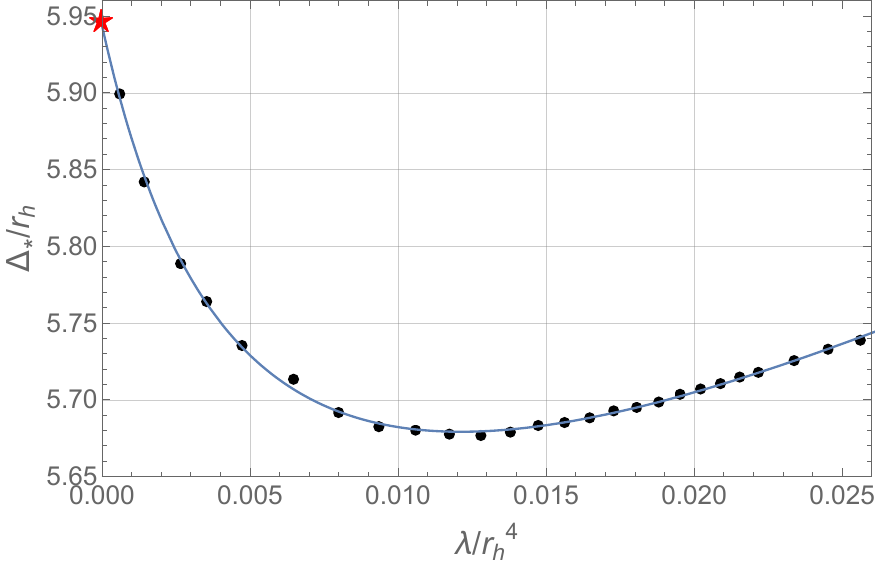}
\caption{Duration of the merger, $\Delta_*/r_h$, as a function of the coupling constant, $\lambda/r_h^4$. The data points were fitted to a Padé approximant with six free parameters. The red star, $\Delta_*=5.94676\, r_h$, corresponds to the value obtained by Emparan and Mart\'inez in \cite{Emparan:2016ylg} for a Schwarzschild black hole ($\lambda/r_h^4=0$). 
\label{fig:deltastar}}
\end{figure}

We can also calculate the duration of the merger, $\Delta_*$, as the retarded time elapsed between the pinch-on instant, $t_*$, and the asymptotic retarded time of the central generator with $q=0$~\cite{Emparan:2016ylg}.
The results are presented in Figure~\ref{fig:deltastar}, and similar to what happened with $r_*$, there is a value of the coupling constant, $\lambda/r_h^4=0.0121$, which minimizes the duration of the merger, in which case $\Delta_*=5.67917\, r_h$. This is not surprising: intuitively, if the small black hole is less distorted, the fused system also takes less time to relax down to a planar horizon. What stands out the most is that the value of $\lambda$ for which $r_*$ is minimized is not the same value that minimizes $\Delta_*$. This discrepancy arises from the interplay between the non-trivial behavior of $r_*$ and $t_*$, both as functions of $\lambda$.
%
%%%%%%%%%
%%%%%%%%%
%

\begin{figure}
\includegraphics[width=8cm]{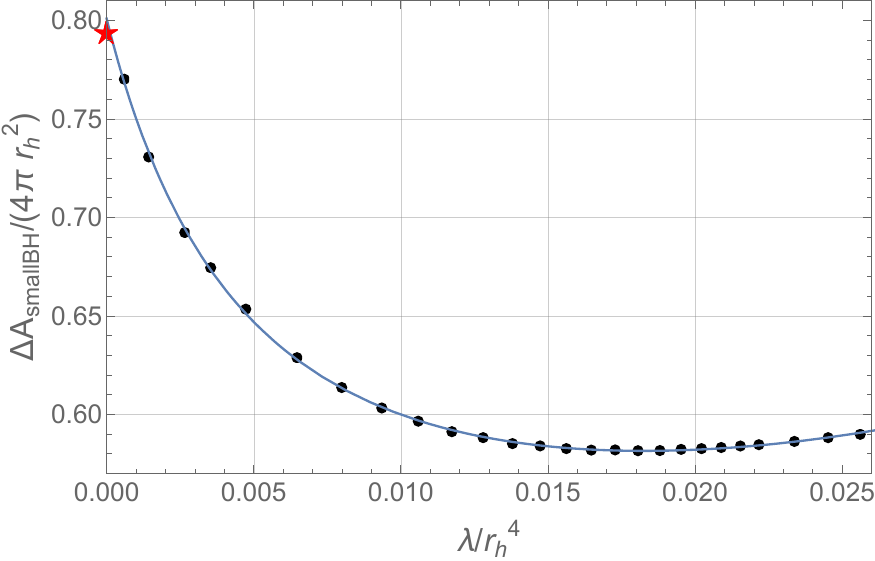}
\caption{Area increment of the small black hole taking into account both non-caustic and caustic generators, $\mathcal{A_\mathrm{smallBH}}/(4\pi r_h^2)$, as a function of the coupling constant, $\lambda/r_h^4$. The data points were fitted to a Padé approximant with six free parameters. The red star, $\Delta \mathcal{A_\mathrm{smallBH}}/(4\pi r_h^2) = 0.79356$, corresponds to the value obtained by Emparan and Mart\'inez in \cite{Emparan:2016ylg} for a Schwarzschild black hole ($\lambda/r_h^4=0$).
\label{fig:area}}
\end{figure}

We may also consider, as in Ref.~\cite{Emparan:2016ylg}, the area growth of the event horizon. At early times the event horizon is disconnected, and the small black hole has the initial area $\mathcal{A}_{in}=4\pi r_h^2$. 
At late times the two initial components of the horizon have fused together and its generators were either already present at the infinite past or they originated from the caustic line.
One can think of the generators with $q>q_*$ as entering the event horizon from the large BH side. Conversely, generators with $q<q_*$ either entered the horizon from the small BH side or else asymptote its apparent horizon in the far past (if $q\leq q_c$).
Considering the contribution of all these generators from the small BH side, they form a disk of area $\pi q_*^2$ at future null infinity, so that, during the process, the area increases by $\Delta \mathcal{A}_{\mathrm{smallBH}}=\left[q_*^2/(4r_h^2)-1\right]\mathcal{A}_{in}$. Figure~\ref{fig:area} illustrates the variation of this quantity with $\lambda$. 
%
%%%%%%%%%
%%%%%%%%%
%
Consistently with the previous results, the area increment of the small black hole also shows a non-monotonic behavior with $\lambda$.

%
%\onecolumngrid\
\begin{table*}[t!]
    \centering
    \begin{tabular}{ccccc}
        \toprule
        & \; Value at $\lambda=0$ \; & \; Value at minimum \; & \; Relative difference \; & \; $\lambda/M^4$ at minimum \; \\ \hline
        \\
      $r_*/r_h$  & 1.75795 (1.76031) & 1.71270 & 2.6\% (2.7\%) & 0.15409\\ 
      $\Delta_*/r_h$   & 5.94409 (5.94676) & 5.67917 & 4.5\% (4.5\%) & 0.36774\\ 
      $\Delta \mathcal{A_\mathrm{smallBH}}/(4\pi r_h^2)$ \; & 0.80090 (0.79356) & 0.58149 & 27.4\% (26.7\%) & 0.71259 \\ \\ \bottomrule
    \end{tabular}
    \caption{Summary of the results for the merger quantities studied: the distortion of the small black hole ($r_*/r_h$), the duration of the merger ($\Delta_*/r_h$) and the area increment of the small black hole ($\Delta \mathcal{A_\mathrm{smallBH}}/(4\pi r_h^2)$). The values outside parenthesis were obtained using the functions fitted to the data, while the values within parenthesis were taken from \cite{Emparan:2016ylg}.}
    \label{tab:my_label}
    \end{table*}
%\twocolumngrid\
%

It is of obvious interest to compare our results with the case of GR, which is recovered in the limit $\lambda\to0$. 
For the Schwarzschild black hole in GR, Emparan and Mart\'inez~\cite{Emparan:2016ylg}  obtained $r_*=1.76031\, r_h$, $\Delta_*=5.94676\, r_h$ and $\Delta \mathcal{A_\mathrm{smallBH}}/(4\pi r_h^2) = 0.79356$. As shown in Figs.~\ref{fig:rstar}, \ref{fig:deltastar} and \ref{fig:area}, we have computed these quantities for 26 different values of $\lambda$, and fitted that data with second-order Pad\'e approximants. The resulting fitting functions extrapolated to $\lambda=0$ match the Emparan-Mart\'inez values with high precision, yielding errors of $0.13\%$, $0.04\%$ and $0.92\%$, respectively.
In other words, we find that in Einsteinian cubic gravity, and for the range of coupling constants studied, the fusion of an EMR binary can happen up to $4.5\%$ quicker than in GR, and the small black hole can be distorted up to $2.7\%$ less, with an area increment that can be up to $26.7\%$ smaller. A summary of these results is presented in Table~\ref{tab:my_label}.

\begin{figure*}[t!]
\centering
\includegraphics[width=0.78\textwidth]{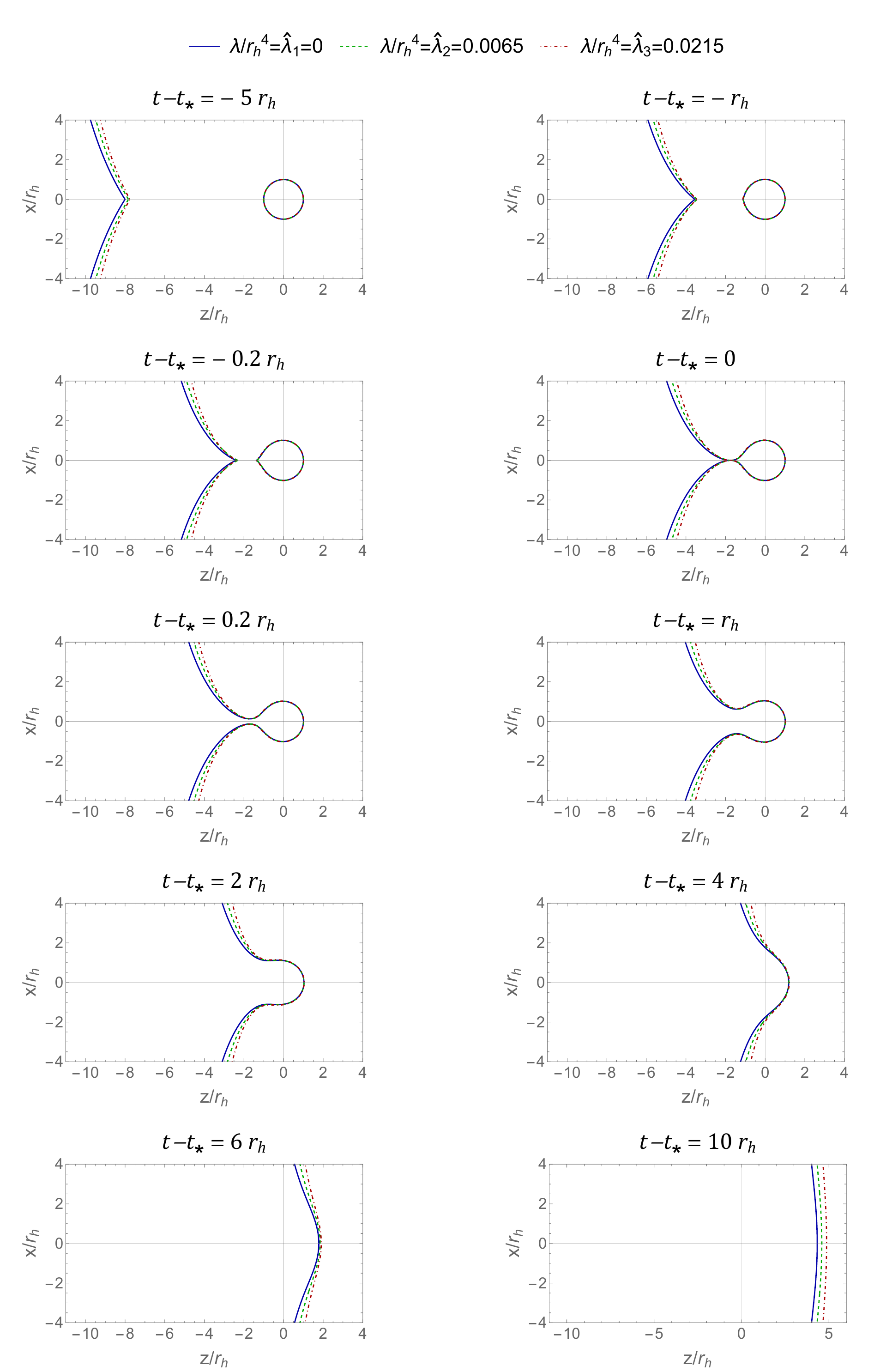}
\caption{Time slices of the event horizon for $\lambda/r_h^4=0$ (blue and thick curves), $\lambda/r_h^4=0.0065$ (green and dashed curves) and $\lambda/r_h^4=0.0215$ (red and dashed-dotted curves), the same choices made for Figs.~\ref{fig:grav_pot} and~\ref{fig:xvsz2}. We used time translation invariance to set $t=t_*$ as the instant at which the small and large BH horizons merge, independently of the value of $\lambda$.
\label{fig:time_slices}}
\end{figure*}

To illustrate how the merger duration depends on the coupling constant, we computed several time slices for the same three $\lambda$ values considered in Fig.~\ref{fig:xvsz2}, overlaying them in Fig.~\ref{fig:time_slices}.

%%%%%%%%%%%%
\subsection{Entropy and area growth}
%%%%%%%%%%%%

In this section, we investigate the connection between the increase in the event horizon area and the entropy growth. Since we have adopted the infinite mass ratio limit, which implies that the area of the large black hole is formally infinite, special care needs to be taken in defining the area increase of the event horizon.

Recall that $\Delta \mathcal{A}_{\mathrm{smallBH}}=\left[q_*^2/(4r_h^2)-1\right]\mathcal{A}_{in}$ is the additional area gained by the small black hole in the merger process, i.e., the difference between the area of the asymptotic disk with radius $q_*$ (see Fig.~\ref{fig:xvsz2}) and the initial area of the small black hole $\mathcal{A}_{in}$.

Accordingly, the total area added to the `large' horizon from generators entering from the small black hole side, is given by
\begin{equation}
\Delta \mathcal{A}_{\mathrm{tot}} = \Delta \mathcal{A}_{\mathrm{smallBH}} + \mathcal{A}_{in}=\left[q_*^2/(4r_h^2)\right]\mathcal{A}_{in} = \pi q_*^2\,.
\end{equation}
Given that the two merging black holes can be regarded initially as two independent systems ---for which the entropy should display the additivity property---, it is natural to compare $\Delta \mathcal{A}_{\mathrm{tot}}$ against the entropy assigned to the small black hole.

In higher curvature theories of gravity, the entropy is computed by Wald's formula~\cite{Wald:1993nt,Iyer:1994ys},
\begin{equation}
S_{\mathrm{Wald}}= - 2\pi \int_H dx^2 \sqrt{h} \frac{\delta {\cal L}}{\delta R_{abcd}} \epsilon_{ab}\epsilon_{cd}\,,
\end{equation}
where the integral is performed over the horizon $H$, with induced metric $h_{ab}$, $h$ being its determinant, and $\epsilon_{ab}$ stands for the volume form of the horizon, satisfying $\epsilon_{ab}\epsilon^{ab} = -2$. 
It turns out that in ECG the Wald entropy of the spherically symmetric BH is known analytically~\cite{Bueno:2016lrh, Hennigar:2016gkm}, even though the spacetime is determined numerically. Specifically it is determined by
\begin{equation}\label{eq:entropy_ecg}
    \frac{S_{\mathrm{Wald}}}{\pi r_h^2}=1-\frac{48\lambda}{r_h^4}\frac{3+2\sqrt{1+48\lambda/r_h^4}}{\left(1+\sqrt{1+48\lambda/r_h^4}\right)^2}+\sqrt{\frac{192\lambda}{r_h^4}}\,,
\end{equation}
normalized so that $S_{\mathrm{Wald}}$ vanishes when $r_h=0$ \footnote{In order to accomplish this, the topological Gauss-Bonnet term must be included in the action. Being topological it does not affect the equations of motion, but it contributes non-trivially to the entropy.}.

\begin{figure}[h!]
\includegraphics[width=8.3cm]{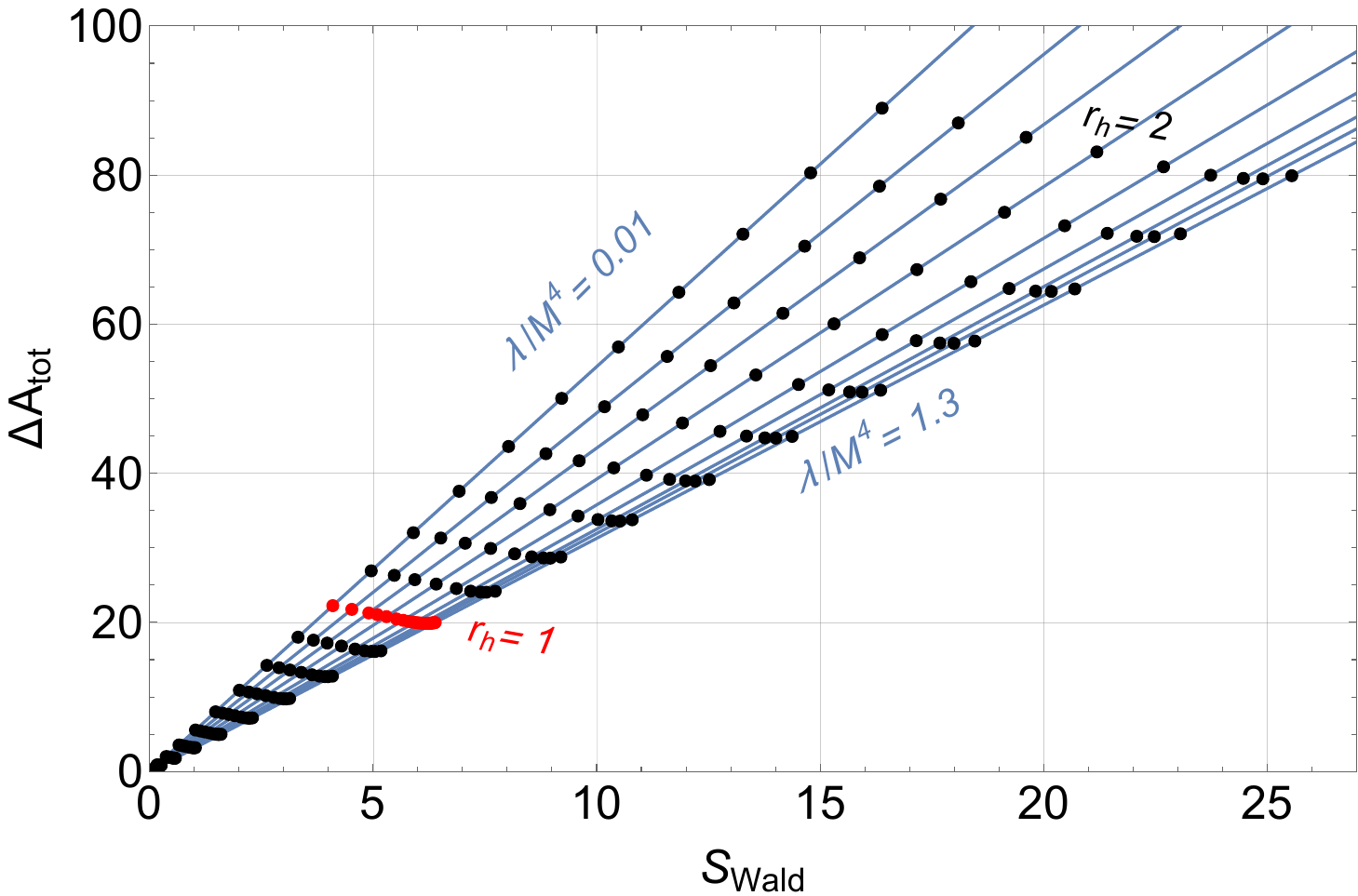}
\caption{Area of the disk formed at future null infinity by the small black hole generators (both non-caustic and caustic), as a function of the Wald entropy of the small black hole, $S_\mathrm{Wald}$. Each family of dots corresponds to a different horizon radius of the small black hole, in the interval $r_h \in [0,2]$, with two adjacent families being spaced by $0.1$. In particular, the red dots correspond to a small black hole with $r_h=1$. Each blue line corresponds to a different value of the dimensionless coupling constant, $\lambda/M^4$ (or $\lambda/r_h^4$).
\label{fig:AvsS}}
\end{figure}

\begin{figure}[t!]
\includegraphics[width=8.8cm]{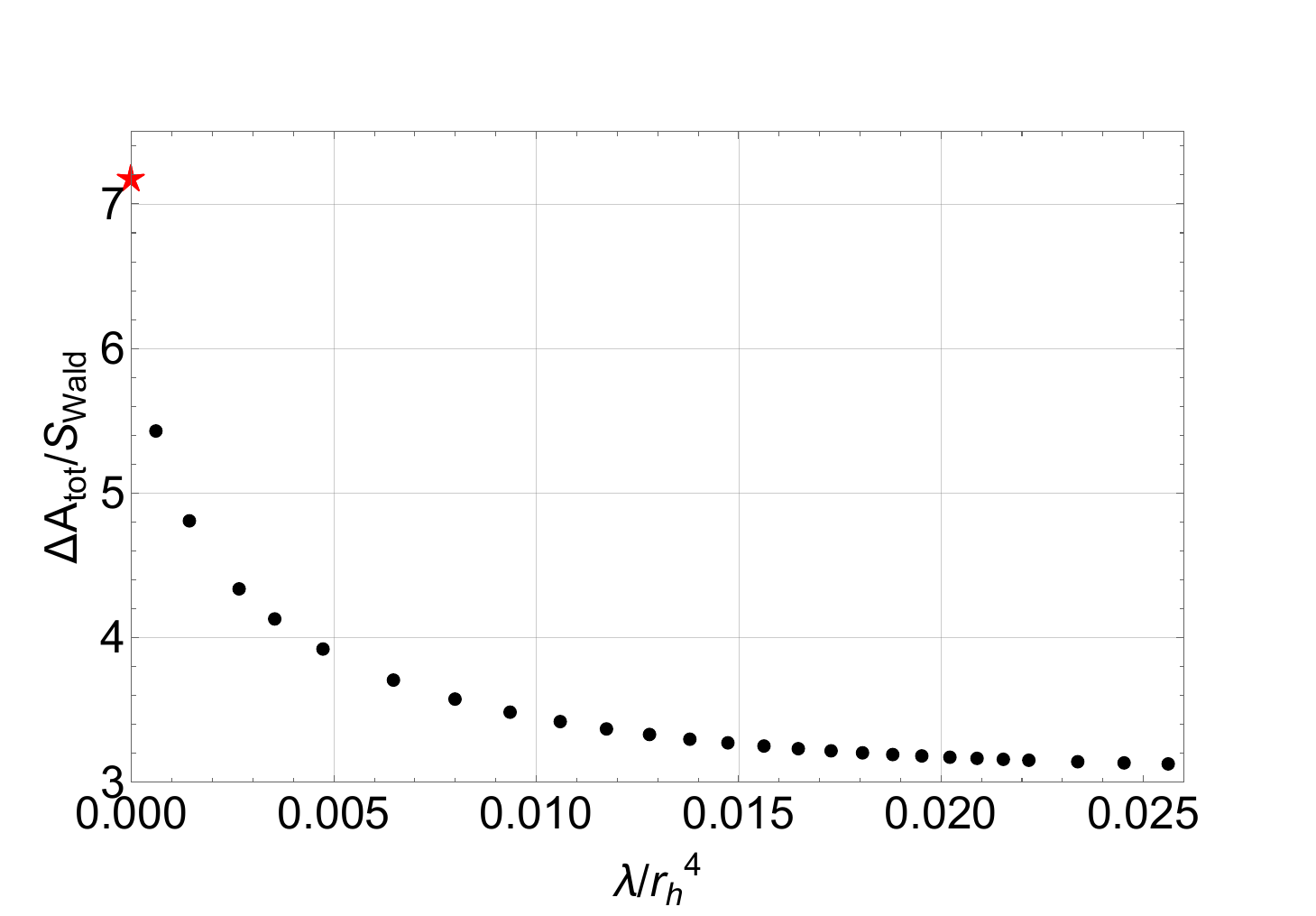}
\caption{Ratio $\Delta \mathcal{A}_{\mathrm{tot}}/S_{\mathrm{Wald}}$ as a function of the dimensionless coupling parameter, $\lambda/r_h^4$.
\label{fig:entropyvsarea2}}
\end{figure}

In Fig.~\ref{fig:AvsS} we show how the Wald entropy of the small black hole, given by Eq.~\eqref{eq:entropy_ecg}, is related to $\Delta \mathcal{A}_{\mathrm{tot}}$.
We do this for different small black hole radii, to emphasize the linear relation between $\Delta \mathcal{A}_{\mathrm{tot}}$ and $S_{\mathrm{Wald}}$, owed to the common dependence in $r_h^2$ (both $\Delta \mathcal{A}_{\mathrm{tot}}/r_h^2$ and $S_{\mathrm{Wald}}/r_h^2$ are uniquely determined by $\lambda/r_h^4$). Hence, on each line of Fig.~\ref{fig:AvsS} we are varying the radius of the black hole, while keeping fixed the coupling parameter $\lambda/r_h^4$ (or $\lambda/M^4$). It is apparent that the slope of these lines decreases with the coupling parameter and, in particular, it seems to approach a constant for large $\lambda/r_h^4$. To better illustrate this, we show in Fig.~\ref{fig:entropyvsarea2} the ratio $\Delta \mathcal{A}_{\mathrm{tot}}/S_{\mathrm{Wald}}$ as a function of the coupling parameter. For $\lambda/M^4 \simeq 1$ we obtain $\Delta \mathcal{A}_{\mathrm{tot}}/S_{\mathrm{Wald}} \simeq 3.12$.

%
%%%%%%%%%
%%%%%%%%%
%

%%%%%%%%%%%%
\section{Discussion and Conclusion\label{sec:Conclusion}}
%%%%%%%%%%%%

In this paper, we have analyzed the merging of two black holes beyond General Relativity, in the extreme-mass ratio regime. The EMR limit allows one to compute the evolution of the coalescing event horizon by null ray-tracing in the background provided by the small black hole. Any extension of GR leading to modifications of the BH geometry, while leaving it asymptotically flat, will produce quantitative changes on the merger process. Since we explore the EMR regime and resolve the scales comparable to the smaller object, this analysis is particularly well-suited to address ultraviolet modifications of GR.

Among the higher derivative theories considered in the literature, we focused on Einsteinian cubic gravity. Besides being astrophysically viable and theoretically well-motivated, this theory only depends on a single coupling constant, $\lambda$, offering therefore a quite manageable parameter space. A previous study~\cite{Hennigar:2018hza} placed an observational constraint on the coupling constant, namely $\lambda < 4.57\times10^{22} M_{\odot}^4$. 
In the present paper, we focused on the effects produced when $\lambda \lesssim M^4$, where $M$ denotes the mass of the small BH. Therefore, taking the small object in our study to be a solar mass BH is well within the viable regime for $\lambda$. On the other hand, considering a supermassive BH would already be in tension with the observational constraint.
The type of analysis conducted herein, when complemented with the associated computation of gravitational wave emission and its confrontation with observations, has the potential to set stronger constraints. To this end it will be instrumental to include corrections beyond the strict EMR limit, to allow the spacetime to be dynamical. This is currently under study.

%
%%%%%%%%%
%%%%%%%%%
%

Our investigation was devoted to a modest range of the dimensionless coupling $\lambda/M^4 \lesssim 1$: this is exactly the regime in which interesting effects are observed.
As the coupling constant is increased starting from zero (i.e., the GR limit), the duration of the merger and the area increment both display non-monotonic behavior: they first decrease to what seems to be an absolute minimum, and from there on they seem to steadily increase. Our numerical results for significantly larger $\lambda$ ---which hinders a precise calculation of the blackening factor--- are inconclusive, leaving uncertain the asymptotic nature of this increase at the moment.
Interestingly, the critical value of $\lambda$ leading to the shortest merging time is {\em different} from the one giving rise to the minimum area increment (see Figs.~\ref{fig:deltastar} and~\ref{fig:area}).

The duration $\Delta_*$ of the horizon fusion gives a measure of how long the merger phase lasts in the GW signal generated from an EMR black hole coalescence. 
Our results mean that there is a specific value of $\lambda/M^4$ for which the merger process is the quickest possible. Compared to GR, at this threshold value of $\lambda$ the merger proceeds in a time $4.5\%$ quicker.

Along the same lines, one is tempted to establish a connection between the increase in the event horizon area and the entropy growth of the binary. Dialing up $\lambda$ starting from zero, the relative change in area initially decreases and note that the difference with respect to GR can be quite significant --- more than $25\%$! Thus, one expects a comparably significant modification in the entropy growth. 
We have seen that the increase of the `large' horizon's area, owed to the generators entering from the small BH side, must be proportional to the Wald entropy of the small BH, for fixed dimensionless coupling $\lambda/r_h^4$. But the constant of proportionality depends strongly on the dimensionless coupling $\lambda/r_h^4$, even for the modest range explored, showing some indications that it might asymptote a constant for large $\lambda/r_h^4$.
Once again, it would be interesting to go beyond the strict EMR limit in order to include gravitational radiation and then use the above ideas to estimate the energy emitted in GWs during such mergers.

In conclusion, we have proposed a means to probe generally covariant theories of gravity (including a class of modifications of GR) in the highly non-linear merger regime. This offers a way to tackle ---with relatively simple analytic techniques--- a problem that typically requires demanding numerical simulations and years-long efforts to adapt perturbative methods to modified gravity. The results presented can serve as a benchmark for upcoming numerical simulations.

\vfill

%
%\onecolumngrid\
%
\pagebreak
\textcolor{white}{BLANK SPACE}

\textcolor{white}{BLANK SPACE}
%\twocolumngrid\
%

%%%%%%%%%%%%%
\begin{acknowledgments}
%%%%%%%%%%%%%
We thank Robie Hennigar for insightful discussions.
AMF and JVR would also like to thank the organizers of the workshop ``Gravity: Challenges beyond General Relativity'', held in Barcelona on May 22-24, 2024, for interesting discussions that took place at that event.
AMF acknowledges the support of the MICINN grants PID2019-105614GB-C22, AGAUR grant 2017-SGR 754, PID2022-136224NB-C22 funded by MCIN/AEI/ 10.13039/501100011033/FEDER, UE, and State Research Agency of MICINN through the ‘Unit of Excellence Maria de Maeztu 2020-2023’ award to the Institute of Cosmos Sciences (CEX2019-000918-M).
DCL and JVR acknowledge support from FCT under Project No. 2022.08368.PTDC. 
JVR also acknowledges support from FCT under Project No. CERN/FIS-PAR/0023/2019.
\end{acknowledgments}

\bibliography{BHmergers_in_ECG_refs}

\begin{thebibliography}{45}%
\makeatletter
\providecommand \@ifxundefined [1]{%
 \@ifx{#1\undefined}
}%
\providecommand \@ifnum [1]{%
 \ifnum #1\expandafter \@firstoftwo
 \else \expandafter \@secondoftwo
 \fi
}%
\providecommand \@ifx [1]{%
 \ifx #1\expandafter \@firstoftwo
 \else \expandafter \@secondoftwo
 \fi
}%
\providecommand \natexlab [1]{#1}%
\providecommand \enquote  [1]{``#1''}%
\providecommand \bibnamefont  [1]{#1}%
\providecommand \bibfnamefont [1]{#1}%
\providecommand \citenamefont [1]{#1}%
\providecommand \href@noop [0]{\@secondoftwo}%
\providecommand \href [0]{\begingroup \@sanitize@url \@href}%
\providecommand \@href[1]{\@@startlink{#1}\@@href}%
\providecommand \@@href[1]{\endgroup#1\@@endlink}%
\providecommand \@sanitize@url [0]{\catcode `\\12\catcode `\$12\catcode
  `\&12\catcode `\#12\catcode `\^12\catcode `\_12\catcode `\%12\relax}%
\providecommand \@@startlink[1]{}%
\providecommand \@@endlink[0]{}%
\providecommand \url  [0]{\begingroup\@sanitize@url \@url }%
\providecommand \@url [1]{\endgroup\@href {#1}{\urlprefix }}%
\providecommand \urlprefix  [0]{URL }%
\providecommand \Eprint [0]{\href }%
\providecommand \doibase [0]{https://doi.org/}%
\providecommand \selectlanguage [0]{\@gobble}%
\providecommand \bibinfo  [0]{\@secondoftwo}%
\providecommand \bibfield  [0]{\@secondoftwo}%
\providecommand \translation [1]{[#1]}%
\providecommand \BibitemOpen [0]{}%
\providecommand \bibitemStop [0]{}%
\providecommand \bibitemNoStop [0]{.\EOS\space}%
\providecommand \EOS [0]{\spacefactor3000\relax}%
\providecommand \BibitemShut  [1]{\csname bibitem#1\endcsname}%
\let\auto@bib@innerbib\@empty
%</preamble>
\bibitem [{\citenamefont {Cervantes-Cota}\ \emph {et~al.}(2016)\citenamefont
  {Cervantes-Cota}, \citenamefont {Galindo-Uribarri},\ and\ \citenamefont
  {Smoot}}]{Cervantes_Cota_2016}%
  \BibitemOpen
  \bibfield  {author} {\bibinfo {author} {\bibfnamefont {J.}~\bibnamefont
  {Cervantes-Cota}}, \bibinfo {author} {\bibfnamefont {S.}~\bibnamefont
  {Galindo-Uribarri}},\ and\ \bibinfo {author} {\bibfnamefont {G.}~\bibnamefont
  {Smoot}},\ }\bibfield  {title} {\bibinfo {title} {A brief history of
  gravitational waves},\ }\href {https://doi.org/10.3390/universe2030022}
  {\bibfield  {journal} {\bibinfo  {journal} {Universe}\ }\textbf {\bibinfo
  {volume} {2}},\ \bibinfo {pages} {22} (\bibinfo {year} {2016})}\BibitemShut
  {NoStop}%
\bibitem [{\citenamefont {Pirani}(1956)}]{Pirani:1956tn}%
  \BibitemOpen
  \bibfield  {author} {\bibinfo {author} {\bibfnamefont {F.~A.~E.}\
  \bibnamefont {Pirani}},\ }\bibfield  {title} {\bibinfo {title} {{On the
  Physical significance of the Riemann tensor}},\ }\href
  {https://doi.org/10.1007/s10714-009-0787-9} {\bibfield  {journal} {\bibinfo
  {journal} {Acta Phys. Polon.}\ }\textbf {\bibinfo {volume} {15}},\ \bibinfo
  {pages} {389} (\bibinfo {year} {1956})}\BibitemShut {NoStop}%
\bibitem [{\citenamefont {Clifton}\ \emph {et~al.}(2012)\citenamefont
  {Clifton}, \citenamefont {Ferreira}, \citenamefont {Padilla},\ and\
  \citenamefont {Skordis}}]{Clifton:2011jh}%
  \BibitemOpen
  \bibfield  {author} {\bibinfo {author} {\bibfnamefont {T.}~\bibnamefont
  {Clifton}}, \bibinfo {author} {\bibfnamefont {P.~G.}\ \bibnamefont
  {Ferreira}}, \bibinfo {author} {\bibfnamefont {A.}~\bibnamefont {Padilla}},\
  and\ \bibinfo {author} {\bibfnamefont {C.}~\bibnamefont {Skordis}},\
  }\bibfield  {title} {\bibinfo {title} {{Modified Gravity and Cosmology}},\
  }\href {https://doi.org/10.1016/j.physrep.2012.01.001} {\bibfield  {journal}
  {\bibinfo  {journal} {Phys. Rept.}\ }\textbf {\bibinfo {volume} {513}},\
  \bibinfo {pages} {1} (\bibinfo {year} {2012})},\ \Eprint
  {https://arxiv.org/abs/1106.2476} {arXiv:1106.2476 [astro-ph.CO]}
  \BibitemShut {NoStop}%
\bibitem [{\citenamefont {Berti}\ \emph {et~al.}(2015)\citenamefont {Berti}
  \emph {et~al.}}]{Berti:2015itd}%
  \BibitemOpen
  \bibfield  {author} {\bibinfo {author} {\bibfnamefont {E.}~\bibnamefont
  {Berti}} \emph {et~al.},\ }\bibfield  {title} {\bibinfo {title} {{Testing
  General Relativity with Present and Future Astrophysical Observations}},\
  }\href {https://doi.org/10.1088/0264-9381/32/24/243001} {\bibfield  {journal}
  {\bibinfo  {journal} {Class. Quant. Grav.}\ }\textbf {\bibinfo {volume}
  {32}},\ \bibinfo {pages} {243001} (\bibinfo {year} {2015})},\ \Eprint
  {https://arxiv.org/abs/1501.07274} {arXiv:1501.07274 [gr-qc]} \BibitemShut
  {NoStop}%
\bibitem [{\citenamefont {Will}(2018)}]{Will:2018bme}%
  \BibitemOpen
  \bibfield  {author} {\bibinfo {author} {\bibfnamefont {C.~M.}\ \bibnamefont
  {Will}},\ }\href@noop {} {\emph {\bibinfo {title} {{Theory and Experiment in
  Gravitational Physics}}}}\ (\bibinfo  {publisher} {Cambridge University
  Press},\ \bibinfo {year} {2018})\BibitemShut {NoStop}%
\bibitem [{\citenamefont {Akrami}\ \emph {et~al.}(2021)\citenamefont {Akrami}
  \emph {et~al.}}]{CANTATA:2021ktz}%
  \BibitemOpen
  \bibfield  {author} {\bibinfo {author} {\bibfnamefont {Y.}~\bibnamefont
  {Akrami}} \emph {et~al.} (\bibinfo {collaboration} {CANTATA}),\ }\href
  {https://doi.org/10.1007/978-3-030-83715-0} {\emph {\bibinfo {title}
  {{Modified Gravity and Cosmology}: {An Update by the CANTATA Network}}}},\
  edited by\ \bibinfo {editor} {\bibfnamefont {E.~N.}\ \bibnamefont
  {Saridakis}}, \bibinfo {editor} {\bibfnamefont {R.}~\bibnamefont {Lazkoz}},
  \bibinfo {editor} {\bibfnamefont {V.}~\bibnamefont {Salzano}}, \bibinfo
  {editor} {\bibfnamefont {P.}~\bibnamefont {Vargas~Moniz}}, \bibinfo {editor}
  {\bibfnamefont {S.}~\bibnamefont {Capozziello}}, \bibinfo {editor}
  {\bibfnamefont {J.}~\bibnamefont {Beltr\'an~Jim\'enez}}, \bibinfo {editor}
  {\bibfnamefont {M.}~\bibnamefont {De~Laurentis}},\ and\ \bibinfo {editor}
  {\bibfnamefont {G.~J.}\ \bibnamefont {Olmo}}\ (\bibinfo  {publisher}
  {Springer},\ \bibinfo {year} {2021})\ \Eprint
  {https://arxiv.org/abs/2105.12582} {arXiv:2105.12582 [gr-qc]} \BibitemShut
  {NoStop}%
\bibitem [{\citenamefont {Yunes}\ and\ \citenamefont
  {Pretorius}(2009)}]{Yunes:2009ke}%
  \BibitemOpen
  \bibfield  {author} {\bibinfo {author} {\bibfnamefont {N.}~\bibnamefont
  {Yunes}}\ and\ \bibinfo {author} {\bibfnamefont {F.}~\bibnamefont
  {Pretorius}},\ }\bibfield  {title} {\bibinfo {title} {{Fundamental
  Theoretical Bias in Gravitational Wave Astrophysics and the Parameterized
  Post-Einsteinian Framework}},\ }\href
  {https://doi.org/10.1103/PhysRevD.80.122003} {\bibfield  {journal} {\bibinfo
  {journal} {Phys. Rev. D}\ }\textbf {\bibinfo {volume} {80}},\ \bibinfo
  {pages} {122003} (\bibinfo {year} {2009})},\ \Eprint
  {https://arxiv.org/abs/0909.3328} {arXiv:0909.3328 [gr-qc]} \BibitemShut
  {NoStop}%
\bibitem [{Note1()}]{Note1}%
  \BibitemOpen
  \bibinfo {note} {For instance, the post-Newtonian approach is commonly used
  to study the inspiral phase, whereas the quasinormal mode analysis is
  well-suited to address the ringdown phase.}\BibitemShut {Stop}%
\bibitem [{\citenamefont {Kov\'acs}\ and\ \citenamefont
  {Reall}(2020)}]{Kovacs:2020ywu}%
  \BibitemOpen
  \bibfield  {author} {\bibinfo {author} {\bibfnamefont {A.~D.}\ \bibnamefont
  {Kov\'acs}}\ and\ \bibinfo {author} {\bibfnamefont {H.~S.}\ \bibnamefont
  {Reall}},\ }\bibfield  {title} {\bibinfo {title} {{Well-posed formulation of
  Lovelock and Horndeski theories}},\ }\href
  {https://doi.org/10.1103/PhysRevD.101.124003} {\bibfield  {journal} {\bibinfo
   {journal} {Phys. Rev. D}\ }\textbf {\bibinfo {volume} {101}},\ \bibinfo
  {pages} {124003} (\bibinfo {year} {2020})},\ \Eprint
  {https://arxiv.org/abs/2003.08398} {arXiv:2003.08398 [gr-qc]} \BibitemShut
  {NoStop}%
\bibitem [{\citenamefont {Barack}\ and\ \citenamefont
  {Pound}(2019)}]{Barack:2018yvs}%
  \BibitemOpen
  \bibfield  {author} {\bibinfo {author} {\bibfnamefont {L.}~\bibnamefont
  {Barack}}\ and\ \bibinfo {author} {\bibfnamefont {A.}~\bibnamefont {Pound}},\
  }\bibfield  {title} {\bibinfo {title} {{Self-force and radiation reaction in
  general relativity}},\ }\href {https://doi.org/10.1088/1361-6633/aae552}
  {\bibfield  {journal} {\bibinfo  {journal} {Rept. Prog. Phys.}\ }\textbf
  {\bibinfo {volume} {82}},\ \bibinfo {pages} {016904} (\bibinfo {year}
  {2019})},\ \Eprint {https://arxiv.org/abs/1805.10385} {arXiv:1805.10385
  [gr-qc]} \BibitemShut {NoStop}%
\bibitem [{\citenamefont {Emparan}\ and\ \citenamefont
  {Mart\'inez}(2016)}]{Emparan:2016ylg}%
  \BibitemOpen
  \bibfield  {author} {\bibinfo {author} {\bibfnamefont {R.}~\bibnamefont
  {Emparan}}\ and\ \bibinfo {author} {\bibfnamefont {M.}~\bibnamefont
  {Mart\'inez}},\ }\bibfield  {title} {\bibinfo {title} {{Exact Event Horizon
  of a Black Hole Merger}},\ }\href
  {https://doi.org/10.1088/0264-9381/33/15/155003} {\bibfield  {journal}
  {\bibinfo  {journal} {Class. Quant. Grav.}\ }\textbf {\bibinfo {volume}
  {33}},\ \bibinfo {pages} {155003} (\bibinfo {year} {2016})},\ \Eprint
  {https://arxiv.org/abs/1603.00712} {arXiv:1603.00712 [gr-qc]} \BibitemShut
  {NoStop}%
\bibitem [{\citenamefont {Colpi}\ \emph {et~al.}(2024)\citenamefont {Colpi}
  \emph {et~al.}}]{Colpi:2024xhw}%
  \BibitemOpen
  \bibfield  {author} {\bibinfo {author} {\bibfnamefont {M.}~\bibnamefont
  {Colpi}} \emph {et~al.},\ }\href@noop {} {\bibinfo {title} {{LISA Definition
  Study Report}}} (\bibinfo {year} {2024}),\ \Eprint
  {https://arxiv.org/abs/2402.07571} {arXiv:2402.07571 [astro-ph.CO]}
  \BibitemShut {NoStop}%
\bibitem [{\citenamefont {Lovelock}(1971)}]{Lovelock:1971yv}%
  \BibitemOpen
  \bibfield  {author} {\bibinfo {author} {\bibfnamefont {D.}~\bibnamefont
  {Lovelock}},\ }\bibfield  {title} {\bibinfo {title} {{The Einstein tensor and
  its generalizations}},\ }\href {https://doi.org/10.1063/1.1665613} {\bibfield
   {journal} {\bibinfo  {journal} {J. Math. Phys.}\ }\textbf {\bibinfo {volume}
  {12}},\ \bibinfo {pages} {498} (\bibinfo {year} {1971})}\BibitemShut
  {NoStop}%
\bibitem [{\citenamefont {Donoghue}(1994)}]{Donoghue:1994dn}%
  \BibitemOpen
  \bibfield  {author} {\bibinfo {author} {\bibfnamefont {J.~F.}\ \bibnamefont
  {Donoghue}},\ }\bibfield  {title} {\bibinfo {title} {{General relativity as
  an effective field theory: The leading quantum corrections}},\ }\href
  {https://doi.org/10.1103/PhysRevD.50.3874} {\bibfield  {journal} {\bibinfo
  {journal} {Phys. Rev. D}\ }\textbf {\bibinfo {volume} {50}},\ \bibinfo
  {pages} {3874} (\bibinfo {year} {1994})},\ \Eprint
  {https://arxiv.org/abs/gr-qc/9405057} {arXiv:gr-qc/9405057} \BibitemShut
  {NoStop}%
\bibitem [{\citenamefont {Stelle}(1977)}]{Stelle:1976gc}%
  \BibitemOpen
  \bibfield  {author} {\bibinfo {author} {\bibfnamefont {K.~S.}\ \bibnamefont
  {Stelle}},\ }\bibfield  {title} {\bibinfo {title} {{Renormalization of Higher
  Derivative Quantum Gravity}},\ }\href
  {https://doi.org/10.1103/PhysRevD.16.953} {\bibfield  {journal} {\bibinfo
  {journal} {Phys. Rev.}\ }\textbf {\bibinfo {volume} {D16}},\ \bibinfo {pages}
  {953} (\bibinfo {year} {1977})}\BibitemShut {NoStop}%
%%CITATION = PHRVA,D16,953;%%
\bibitem [{\citenamefont {De~Felice}\ and\ \citenamefont
  {Tsujikawa}(2010)}]{DeFelice:2010aj}%
  \BibitemOpen
  \bibfield  {author} {\bibinfo {author} {\bibfnamefont {A.}~\bibnamefont
  {De~Felice}}\ and\ \bibinfo {author} {\bibfnamefont {S.}~\bibnamefont
  {Tsujikawa}},\ }\bibfield  {title} {\bibinfo {title} {{f(R) theories}},\
  }\href {https://doi.org/10.12942/lrr-2010-3} {\bibfield  {journal} {\bibinfo
  {journal} {Living Rev. Rel.}\ }\textbf {\bibinfo {volume} {13}},\ \bibinfo
  {pages} {3} (\bibinfo {year} {2010})},\ \Eprint
  {https://arxiv.org/abs/1002.4928} {arXiv:1002.4928 [gr-qc]} \BibitemShut
  {NoStop}%
%%CITATION = ARXIV:1002.4928;%%
\bibitem [{\citenamefont {Holdom}(2002)}]{Holdom:2002xy}%
  \BibitemOpen
  \bibfield  {author} {\bibinfo {author} {\bibfnamefont {B.}~\bibnamefont
  {Holdom}},\ }\bibfield  {title} {\bibinfo {title} {{On the fate of
  singularities and horizons in higher derivative gravity}},\ }\href
  {https://doi.org/10.1103/PhysRevD.66.084010} {\bibfield  {journal} {\bibinfo
  {journal} {Phys. Rev.}\ }\textbf {\bibinfo {volume} {D66}},\ \bibinfo {pages}
  {084010} (\bibinfo {year} {2002})},\ \Eprint
  {https://arxiv.org/abs/hep-th/0206219} {arXiv:hep-th/0206219 [hep-th]}
  \BibitemShut {NoStop}%
%%CITATION = HEP-TH/0206219;%%
\bibitem [{\citenamefont {Mignemi}\ and\ \citenamefont
  {Wiltshire}(1992)}]{Mignemi:1991wa}%
  \BibitemOpen
  \bibfield  {author} {\bibinfo {author} {\bibfnamefont {S.}~\bibnamefont
  {Mignemi}}\ and\ \bibinfo {author} {\bibfnamefont {D.~L.}\ \bibnamefont
  {Wiltshire}},\ }\bibfield  {title} {\bibinfo {title} {{Black holes in higher
  derivative gravity theories}},\ }\href
  {https://doi.org/10.1103/PhysRevD.46.1475} {\bibfield  {journal} {\bibinfo
  {journal} {Phys. Rev.}\ }\textbf {\bibinfo {volume} {D46}},\ \bibinfo {pages}
  {1475} (\bibinfo {year} {1992})},\ \Eprint
  {https://arxiv.org/abs/hep-th/9202031} {arXiv:hep-th/9202031 [hep-th]}
  \BibitemShut {NoStop}%
%%CITATION = HEP-TH/9202031;%%
\bibitem [{\citenamefont {Nelson}(2010)}]{Nelson:2010ig}%
  \BibitemOpen
  \bibfield  {author} {\bibinfo {author} {\bibfnamefont {W.}~\bibnamefont
  {Nelson}},\ }\bibfield  {title} {\bibinfo {title} {{Static Solutions for 4th
  order gravity}},\ }\href {https://doi.org/10.1103/PhysRevD.82.104026}
  {\bibfield  {journal} {\bibinfo  {journal} {Phys. Rev.}\ }\textbf {\bibinfo
  {volume} {D82}},\ \bibinfo {pages} {104026} (\bibinfo {year} {2010})},\
  \Eprint {https://arxiv.org/abs/1010.3986} {arXiv:1010.3986 [gr-qc]}
  \BibitemShut {NoStop}%
%%CITATION = ARXIV:1010.3986;%%
\bibitem [{\citenamefont {Bueno}\ and\ \citenamefont
  {Cano}(2017)}]{Bueno:2017sui}%
  \BibitemOpen
  \bibfield  {author} {\bibinfo {author} {\bibfnamefont {P.}~\bibnamefont
  {Bueno}}\ and\ \bibinfo {author} {\bibfnamefont {P.~A.}\ \bibnamefont
  {Cano}},\ }\bibfield  {title} {\bibinfo {title} {{On black holes in
  higher-derivative gravities}},\ }\href
  {https://doi.org/10.1088/1361-6382/aa8056} {\bibfield  {journal} {\bibinfo
  {journal} {Class. Quant. Grav.}\ }\textbf {\bibinfo {volume} {34}},\ \bibinfo
  {pages} {175008} (\bibinfo {year} {2017})},\ \Eprint
  {https://arxiv.org/abs/1703.04625} {arXiv:1703.04625 [hep-th]} \BibitemShut
  {NoStop}%
%%CITATION = ARXIV:1703.04625;%%
\bibitem [{\citenamefont {Goldstein}\ and\ \citenamefont
  {Mashiyane}(2018)}]{Goldstein:2017rxn}%
  \BibitemOpen
  \bibfield  {author} {\bibinfo {author} {\bibfnamefont {K.}~\bibnamefont
  {Goldstein}}\ and\ \bibinfo {author} {\bibfnamefont {J.~J.}\ \bibnamefont
  {Mashiyane}},\ }\bibfield  {title} {\bibinfo {title} {{Ineffective Higher
  Derivative Black Hole Hair}},\ }\href
  {https://doi.org/10.1103/PhysRevD.97.024015} {\bibfield  {journal} {\bibinfo
  {journal} {Phys. Rev.}\ }\textbf {\bibinfo {volume} {D97}},\ \bibinfo {pages}
  {024015} (\bibinfo {year} {2018})},\ \Eprint
  {https://arxiv.org/abs/1703.02803} {arXiv:1703.02803 [hep-th]} \BibitemShut
  {NoStop}%
%%CITATION = ARXIV:1703.02803;%%
\bibitem [{\citenamefont {Myers}\ and\ \citenamefont
  {Robinson}(2010)}]{Myers:2010ru}%
  \BibitemOpen
  \bibfield  {author} {\bibinfo {author} {\bibfnamefont {R.~C.}\ \bibnamefont
  {Myers}}\ and\ \bibinfo {author} {\bibfnamefont {B.}~\bibnamefont
  {Robinson}},\ }\bibfield  {title} {\bibinfo {title} {{Black Holes in
  Quasi-topological Gravity}},\ }\href
  {https://doi.org/10.1007/JHEP08(2010)067} {\bibfield  {journal} {\bibinfo
  {journal} {JHEP}\ }\textbf {\bibinfo {volume} {08}},\ \bibinfo {pages}
  {067}},\ \Eprint {https://arxiv.org/abs/1003.5357} {arXiv:1003.5357 [gr-qc]}
  \BibitemShut {NoStop}%
%%CITATION = ARXIV:1003.5357;%%
\bibitem [{\citenamefont {Myers}\ \emph {et~al.}(2010)\citenamefont {Myers},
  \citenamefont {Paulos},\ and\ \citenamefont {Sinha}}]{Myers:2010jv}%
  \BibitemOpen
  \bibfield  {author} {\bibinfo {author} {\bibfnamefont {R.~C.}\ \bibnamefont
  {Myers}}, \bibinfo {author} {\bibfnamefont {M.~F.}\ \bibnamefont {Paulos}},\
  and\ \bibinfo {author} {\bibfnamefont {A.}~\bibnamefont {Sinha}},\ }\bibfield
   {title} {\bibinfo {title} {{Holographic studies of quasi-topological
  gravity}},\ }\href {https://doi.org/10.1007/JHEP08(2010)035} {\bibfield
  {journal} {\bibinfo  {journal} {JHEP}\ }\textbf {\bibinfo {volume} {08}},\
  \bibinfo {pages} {035}},\ \Eprint {https://arxiv.org/abs/1004.2055}
  {arXiv:1004.2055 [hep-th]} \BibitemShut {NoStop}%
%%CITATION = ARXIV:1004.2055;%%
\bibitem [{\citenamefont {L{\"u}}\ \emph
  {et~al.}(2015{\natexlab{a}})\citenamefont {L{\"u}}, \citenamefont {Perkins},
  \citenamefont {Pope},\ and\ \citenamefont {Stelle}}]{Lu:2015cqa}%
  \BibitemOpen
  \bibfield  {author} {\bibinfo {author} {\bibfnamefont {H.}~\bibnamefont
  {L{\"u}}}, \bibinfo {author} {\bibfnamefont {A.}~\bibnamefont {Perkins}},
  \bibinfo {author} {\bibfnamefont {C.~N.}\ \bibnamefont {Pope}},\ and\
  \bibinfo {author} {\bibfnamefont {K.~S.}\ \bibnamefont {Stelle}},\ }\bibfield
   {title} {\bibinfo {title} {{Black Holes in Higher-Derivative Gravity}},\
  }\href {https://doi.org/10.1103/PhysRevLett.114.171601} {\bibfield  {journal}
  {\bibinfo  {journal} {Phys. Rev. Lett.}\ }\textbf {\bibinfo {volume} {114}},\
  \bibinfo {pages} {171601} (\bibinfo {year} {2015}{\natexlab{a}})},\ \Eprint
  {https://arxiv.org/abs/1502.01028} {arXiv:1502.01028 [hep-th]} \BibitemShut
  {NoStop}%
%%CITATION = ARXIV:1502.01028;%%
\bibitem [{\citenamefont {L{\"u}}\ \emph
  {et~al.}(2015{\natexlab{b}})\citenamefont {L{\"u}}, \citenamefont {Perkins},
  \citenamefont {Pope},\ and\ \citenamefont {Stelle}}]{Lu:2015psa}%
  \BibitemOpen
  \bibfield  {author} {\bibinfo {author} {\bibfnamefont {H.}~\bibnamefont
  {L{\"u}}}, \bibinfo {author} {\bibfnamefont {A.}~\bibnamefont {Perkins}},
  \bibinfo {author} {\bibfnamefont {C.~N.}\ \bibnamefont {Pope}},\ and\
  \bibinfo {author} {\bibfnamefont {K.~S.}\ \bibnamefont {Stelle}},\ }\bibfield
   {title} {\bibinfo {title} {{Spherically Symmetric Solutions in
  Higher-Derivative Gravity}},\ }\href
  {https://doi.org/10.1103/PhysRevD.92.124019} {\bibfield  {journal} {\bibinfo
  {journal} {Phys. Rev.}\ }\textbf {\bibinfo {volume} {D92}},\ \bibinfo {pages}
  {124019} (\bibinfo {year} {2015}{\natexlab{b}})},\ \Eprint
  {https://arxiv.org/abs/1508.00010} {arXiv:1508.00010 [hep-th]} \BibitemShut
  {NoStop}%
%%CITATION = ARXIV:1508.00010;%%
\bibitem [{\citenamefont {Bueno}\ and\ \citenamefont
  {Cano}(2016{\natexlab{a}})}]{Bueno:2016xff}%
  \BibitemOpen
  \bibfield  {author} {\bibinfo {author} {\bibfnamefont {P.}~\bibnamefont
  {Bueno}}\ and\ \bibinfo {author} {\bibfnamefont {P.~A.}\ \bibnamefont
  {Cano}},\ }\bibfield  {title} {\bibinfo {title} {{Einsteinian cubic
  gravity}},\ }\href {https://doi.org/10.1103/PhysRevD.94.104005} {\bibfield
  {journal} {\bibinfo  {journal} {Phys. Rev.}\ }\textbf {\bibinfo {volume}
  {D94}},\ \bibinfo {pages} {104005} (\bibinfo {year} {2016}{\natexlab{a}})},\
  \Eprint {https://arxiv.org/abs/1607.06463} {arXiv:1607.06463 [hep-th]}
  \BibitemShut {NoStop}%
%%CITATION = ARXIV:1607.06463;%%
\bibitem [{\citenamefont {Hennigar}\ \emph {et~al.}(2017)\citenamefont
  {Hennigar}, \citenamefont {Kubiz\v{n}\'ak},\ and\ \citenamefont
  {Mann}}]{Hennigar:2017ego}%
  \BibitemOpen
  \bibfield  {author} {\bibinfo {author} {\bibfnamefont {R.~A.}\ \bibnamefont
  {Hennigar}}, \bibinfo {author} {\bibfnamefont {D.}~\bibnamefont
  {Kubiz\v{n}\'ak}},\ and\ \bibinfo {author} {\bibfnamefont {R.~B.}\
  \bibnamefont {Mann}},\ }\bibfield  {title} {\bibinfo {title} {{Generalized
  quasitopological gravity}},\ }\href
  {https://doi.org/10.1103/PhysRevD.95.104042} {\bibfield  {journal} {\bibinfo
  {journal} {Phys. Rev.}\ }\textbf {\bibinfo {volume} {D95}},\ \bibinfo {pages}
  {104042} (\bibinfo {year} {2017})},\ \Eprint
  {https://arxiv.org/abs/1703.01631} {arXiv:1703.01631 [hep-th]} \BibitemShut
  {NoStop}%
%%CITATION = ARXIV:1703.01631;%%
\bibitem [{\citenamefont {Arciniega}\ \emph
  {et~al.}(2020{\natexlab{a}})\citenamefont {Arciniega}, \citenamefont
  {Edelstein},\ and\ \citenamefont {Jaime}}]{Arciniega:2018fxj}%
  \BibitemOpen
  \bibfield  {author} {\bibinfo {author} {\bibfnamefont {G.}~\bibnamefont
  {Arciniega}}, \bibinfo {author} {\bibfnamefont {J.~D.}\ \bibnamefont
  {Edelstein}},\ and\ \bibinfo {author} {\bibfnamefont {L.~G.}\ \bibnamefont
  {Jaime}},\ }\bibfield  {title} {\bibinfo {title} {{Towards geometric
  inflation: the cubic case}},\ }\href
  {https://doi.org/10.1016/j.physletb.2020.135272} {\bibfield  {journal}
  {\bibinfo  {journal} {Phys. Lett. B}\ }\textbf {\bibinfo {volume} {802}},\
  \bibinfo {pages} {135272} (\bibinfo {year} {2020}{\natexlab{a}})},\ \Eprint
  {https://arxiv.org/abs/1810.08166} {arXiv:1810.08166 [gr-qc]} \BibitemShut
  {NoStop}%
\bibitem [{\citenamefont {Arciniega}\ \emph
  {et~al.}(2020{\natexlab{b}})\citenamefont {Arciniega}, \citenamefont {Bueno},
  \citenamefont {Cano}, \citenamefont {Edelstein}, \citenamefont {Hennigar},\
  and\ \citenamefont {Jaime}}]{Arciniega:2018tnn}%
  \BibitemOpen
  \bibfield  {author} {\bibinfo {author} {\bibfnamefont {G.}~\bibnamefont
  {Arciniega}}, \bibinfo {author} {\bibfnamefont {P.}~\bibnamefont {Bueno}},
  \bibinfo {author} {\bibfnamefont {P.~A.}\ \bibnamefont {Cano}}, \bibinfo
  {author} {\bibfnamefont {J.~D.}\ \bibnamefont {Edelstein}}, \bibinfo {author}
  {\bibfnamefont {R.~A.}\ \bibnamefont {Hennigar}},\ and\ \bibinfo {author}
  {\bibfnamefont {L.~G.}\ \bibnamefont {Jaime}},\ }\bibfield  {title} {\bibinfo
  {title} {{Geometric Inflation}},\ }\href
  {https://doi.org/10.1016/j.physletb.2020.135242} {\bibfield  {journal}
  {\bibinfo  {journal} {Phys. Lett. B}\ }\textbf {\bibinfo {volume} {802}},\
  \bibinfo {pages} {135242} (\bibinfo {year} {2020}{\natexlab{b}})},\ \Eprint
  {https://arxiv.org/abs/1812.11187} {arXiv:1812.11187 [hep-th]} \BibitemShut
  {NoStop}%
\bibitem [{\citenamefont {Hennigar}\ \emph {et~al.}(2018)\citenamefont
  {Hennigar}, \citenamefont {Poshteh},\ and\ \citenamefont
  {Mann}}]{Hennigar:2018hza}%
  \BibitemOpen
  \bibfield  {author} {\bibinfo {author} {\bibfnamefont {R.~A.}\ \bibnamefont
  {Hennigar}}, \bibinfo {author} {\bibfnamefont {M.~B.~J.}\ \bibnamefont
  {Poshteh}},\ and\ \bibinfo {author} {\bibfnamefont {R.~B.}\ \bibnamefont
  {Mann}},\ }\bibfield  {title} {\bibinfo {title} {{Shadows, Signals, and
  Stability in Einsteinian Cubic Gravity}},\ }\href
  {https://doi.org/10.1103/PhysRevD.97.064041} {\bibfield  {journal} {\bibinfo
  {journal} {Phys. Rev.}\ }\textbf {\bibinfo {volume} {D97}},\ \bibinfo {pages}
  {064041} (\bibinfo {year} {2018})},\ \Eprint
  {https://arxiv.org/abs/1801.03223} {arXiv:1801.03223 [gr-qc]} \BibitemShut
  {NoStop}%
%%CITATION = ARXIV:1801.03223;%%
\bibitem [{\citenamefont {Poshteh}\ and\ \citenamefont
  {Mann}(2019)}]{Poshteh:2018wqy}%
  \BibitemOpen
  \bibfield  {author} {\bibinfo {author} {\bibfnamefont {M.~B.~J.}\
  \bibnamefont {Poshteh}}\ and\ \bibinfo {author} {\bibfnamefont {R.~B.}\
  \bibnamefont {Mann}},\ }\bibfield  {title} {\bibinfo {title} {{Gravitational
  Lensing by Black Holes in Einsteinian Cubic Gravity}},\ }\href
  {https://doi.org/10.1103/PhysRevD.99.024035} {\bibfield  {journal} {\bibinfo
  {journal} {Phys. Rev.}\ }\textbf {\bibinfo {volume} {D99}},\ \bibinfo {pages}
  {024035} (\bibinfo {year} {2019})},\ \Eprint
  {https://arxiv.org/abs/1810.10657} {arXiv:1810.10657 [gr-qc]} \BibitemShut
  {NoStop}%
%%CITATION = ARXIV:1810.10657;%%
\bibitem [{\citenamefont {Bueno}\ and\ \citenamefont
  {Cano}(2016{\natexlab{b}})}]{Bueno:2016lrh}%
  \BibitemOpen
  \bibfield  {author} {\bibinfo {author} {\bibfnamefont {P.}~\bibnamefont
  {Bueno}}\ and\ \bibinfo {author} {\bibfnamefont {P.~A.}\ \bibnamefont
  {Cano}},\ }\bibfield  {title} {\bibinfo {title} {{Four-dimensional black
  holes in Einsteinian cubic gravity}},\ }\href
  {https://doi.org/10.1103/PhysRevD.94.124051} {\bibfield  {journal} {\bibinfo
  {journal} {Phys. Rev.}\ }\textbf {\bibinfo {volume} {D94}},\ \bibinfo {pages}
  {124051} (\bibinfo {year} {2016}{\natexlab{b}})},\ \Eprint
  {https://arxiv.org/abs/1610.08019} {arXiv:1610.08019 [hep-th]} \BibitemShut
  {NoStop}%
%%CITATION = ARXIV:1610.08019;%%
\bibitem [{\citenamefont {Hennigar}\ and\ \citenamefont
  {Mann}(2017)}]{Hennigar:2016gkm}%
  \BibitemOpen
  \bibfield  {author} {\bibinfo {author} {\bibfnamefont {R.~A.}\ \bibnamefont
  {Hennigar}}\ and\ \bibinfo {author} {\bibfnamefont {R.~B.}\ \bibnamefont
  {Mann}},\ }\bibfield  {title} {\bibinfo {title} {{Black holes in Einsteinian
  cubic gravity}},\ }\href {https://doi.org/10.1103/PhysRevD.95.064055}
  {\bibfield  {journal} {\bibinfo  {journal} {Phys. Rev.}\ }\textbf {\bibinfo
  {volume} {D95}},\ \bibinfo {pages} {064055} (\bibinfo {year} {2017})},\
  \Eprint {https://arxiv.org/abs/1610.06675} {arXiv:1610.06675 [hep-th]}
  \BibitemShut {NoStop}%
%%CITATION = ARXIV:1610.06675;%%
\bibitem [{\citenamefont {Frassino}\ and\ \citenamefont
  {Rocha}(2020)}]{Frassino:2020zuv}%
  \BibitemOpen
  \bibfield  {author} {\bibinfo {author} {\bibfnamefont {A.~M.}\ \bibnamefont
  {Frassino}}\ and\ \bibinfo {author} {\bibfnamefont {J.~V.}\ \bibnamefont
  {Rocha}},\ }\bibfield  {title} {\bibinfo {title} {{Charged black holes in
  Einsteinian cubic gravity and nonuniqueness}},\ }\href
  {https://doi.org/10.1103/PhysRevD.102.024035} {\bibfield  {journal} {\bibinfo
   {journal} {Phys. Rev. D}\ }\textbf {\bibinfo {volume} {102}},\ \bibinfo
  {pages} {024035} (\bibinfo {year} {2020})},\ \Eprint
  {https://arxiv.org/abs/2002.04071} {arXiv:2002.04071 [hep-th]} \BibitemShut
  {NoStop}%
\bibitem [{\citenamefont {de~Rham}\ \emph {et~al.}(2020)\citenamefont
  {de~Rham}, \citenamefont {Francfort},\ and\ \citenamefont
  {Zhang}}]{deRham:2020ejn}%
  \BibitemOpen
  \bibfield  {author} {\bibinfo {author} {\bibfnamefont {C.}~\bibnamefont
  {de~Rham}}, \bibinfo {author} {\bibfnamefont {J.}~\bibnamefont {Francfort}},\
  and\ \bibinfo {author} {\bibfnamefont {J.}~\bibnamefont {Zhang}},\ }\bibfield
   {title} {\bibinfo {title} {{Black Hole Gravitational Waves in the Effective
  Field Theory of Gravity}},\ }\href
  {https://doi.org/10.1103/PhysRevD.102.024079} {\bibfield  {journal} {\bibinfo
   {journal} {Phys. Rev. D}\ }\textbf {\bibinfo {volume} {102}},\ \bibinfo
  {pages} {024079} (\bibinfo {year} {2020})},\ \Eprint
  {https://arxiv.org/abs/2005.13923} {arXiv:2005.13923 [hep-th]} \BibitemShut
  {NoStop}%
\bibitem [{\citenamefont {Emparan}\ \emph {et~al.}(2018)\citenamefont
  {Emparan}, \citenamefont {Mart\'inez},\ and\ \citenamefont
  {Zilh\~ao}}]{Emparan:2017vyp}%
  \BibitemOpen
  \bibfield  {author} {\bibinfo {author} {\bibfnamefont {R.}~\bibnamefont
  {Emparan}}, \bibinfo {author} {\bibfnamefont {M.}~\bibnamefont
  {Mart\'inez}},\ and\ \bibinfo {author} {\bibfnamefont {M.}~\bibnamefont
  {Zilh\~ao}},\ }\bibfield  {title} {\bibinfo {title} {{Black hole fusion in
  the extreme mass ratio limit}},\ }\href
  {https://doi.org/10.1103/PhysRevD.97.044004} {\bibfield  {journal} {\bibinfo
  {journal} {Phys. Rev. D}\ }\textbf {\bibinfo {volume} {97}},\ \bibinfo
  {pages} {044004} (\bibinfo {year} {2018})},\ \Eprint
  {https://arxiv.org/abs/1708.08868} {arXiv:1708.08868 [gr-qc]} \BibitemShut
  {NoStop}%
\bibitem [{\citenamefont {Emparan}\ and\ \citenamefont
  {Mar\'in}(2020)}]{Emparan:2020uvt}%
  \BibitemOpen
  \bibfield  {author} {\bibinfo {author} {\bibfnamefont {R.}~\bibnamefont
  {Emparan}}\ and\ \bibinfo {author} {\bibfnamefont {D.}~\bibnamefont
  {Mar\'in}},\ }\bibfield  {title} {\bibinfo {title} {{Precursory collapse in
  Neutron Star-Black Hole mergers}},\ }\href
  {https://doi.org/10.1103/PhysRevD.102.024009} {\bibfield  {journal} {\bibinfo
   {journal} {Phys. Rev. D}\ }\textbf {\bibinfo {volume} {102}},\ \bibinfo
  {pages} {024009} (\bibinfo {year} {2020})},\ \Eprint
  {https://arxiv.org/abs/2004.08143} {arXiv:2004.08143 [gr-qc]} \BibitemShut
  {NoStop}%
\bibitem [{\citenamefont {Pina}\ \emph {et~al.}(2022)\citenamefont {Pina},
  \citenamefont {Orselli},\ and\ \citenamefont {Pica}}]{Pina:2022dye}%
  \BibitemOpen
  \bibfield  {author} {\bibinfo {author} {\bibfnamefont {D.~M.}\ \bibnamefont
  {Pina}}, \bibinfo {author} {\bibfnamefont {M.}~\bibnamefont {Orselli}},\ and\
  \bibinfo {author} {\bibfnamefont {D.}~\bibnamefont {Pica}},\ }\bibfield
  {title} {\bibinfo {title} {{Event horizon of a charged black hole binary
  merger}},\ }\href {https://doi.org/10.1103/PhysRevD.106.084012} {\bibfield
  {journal} {\bibinfo  {journal} {Phys. Rev. D}\ }\textbf {\bibinfo {volume}
  {106}},\ \bibinfo {pages} {084012} (\bibinfo {year} {2022})},\ \Eprint
  {https://arxiv.org/abs/2204.08841} {arXiv:2204.08841 [gr-qc]} \BibitemShut
  {NoStop}%
\bibitem [{\citenamefont {Dias}\ \emph {et~al.}(2023)\citenamefont {Dias},
  \citenamefont {Frassino}, \citenamefont {Paccoia},\ and\ \citenamefont
  {Rocha}}]{Dias:2023pdx}%
  \BibitemOpen
  \bibfield  {author} {\bibinfo {author} {\bibfnamefont {J.~M.}\ \bibnamefont
  {Dias}}, \bibinfo {author} {\bibfnamefont {A.~M.}\ \bibnamefont {Frassino}},
  \bibinfo {author} {\bibfnamefont {V.~D.}\ \bibnamefont {Paccoia}},\ and\
  \bibinfo {author} {\bibfnamefont {J.~V.}\ \bibnamefont {Rocha}},\ }\bibfield
  {title} {\bibinfo {title} {{Black hole-wormhole collisions and the emergence
  of islands}},\ }\href {https://doi.org/10.1103/PhysRevD.107.124056}
  {\bibfield  {journal} {\bibinfo  {journal} {Phys. Rev. D}\ }\textbf {\bibinfo
  {volume} {107}},\ \bibinfo {pages} {124056} (\bibinfo {year} {2023})},\
  \Eprint {https://arxiv.org/abs/2304.06098} {arXiv:2304.06098 [gr-qc]}
  \BibitemShut {NoStop}%
\bibitem [{Note2()}]{Note2}%
  \BibitemOpen
  \bibinfo {note} {Note that $r_s/r_h$ tends to be larger for larger values of
  $\lambda /r_h^4$, and so the numerical solution is valid in a wider range as
  we increase the coupling parameter. This is because, as we increase $\lambda
  /r_h^4$, we decrease the exponent of the asymptotic behavior $\sim \protect
  \qopname \relax o{exp}\left (\protect \frac {(r/r_h)^{5/2}}{\lambda
  /r_h^4}\right )$, which delays its contribution to the numerical
  solution.}\BibitemShut {Stop}%
\bibitem [{\citenamefont {Gadioux}\ and\ \citenamefont
  {Reall}(2023)}]{Gadioux:2023pmw}%
  \BibitemOpen
  \bibfield  {author} {\bibinfo {author} {\bibfnamefont {M.}~\bibnamefont
  {Gadioux}}\ and\ \bibinfo {author} {\bibfnamefont {H.~S.}\ \bibnamefont
  {Reall}},\ }\bibfield  {title} {\bibinfo {title} {{Creases, corners, and
  caustics: Properties of nonsmooth structures on black hole horizons}},\
  }\href {https://doi.org/10.1103/PhysRevD.108.084021} {\bibfield  {journal}
  {\bibinfo  {journal} {Phys. Rev. D}\ }\textbf {\bibinfo {volume} {108}},\
  \bibinfo {pages} {084021} (\bibinfo {year} {2023})},\ \Eprint
  {https://arxiv.org/abs/2303.15512} {arXiv:2303.15512 [gr-qc]} \BibitemShut
  {NoStop}%
\bibitem [{\citenamefont {Gadioux}\ \emph {et~al.}(2024)\citenamefont
  {Gadioux}, \citenamefont {Hennigar},\ and\ \citenamefont
  {Reall}}]{Gadioux:2024tlm}%
  \BibitemOpen
  \bibfield  {author} {\bibinfo {author} {\bibfnamefont {M.}~\bibnamefont
  {Gadioux}}, \bibinfo {author} {\bibfnamefont {R.~A.}\ \bibnamefont
  {Hennigar}},\ and\ \bibinfo {author} {\bibfnamefont {H.~S.}\ \bibnamefont
  {Reall}},\ }\href@noop {} {\bibinfo {title} {{Evolution of creases on the
  event horizon of a black hole merger}}} (\bibinfo {year} {2024}),\ \Eprint
  {https://arxiv.org/abs/2407.07962} {arXiv:2407.07962 [gr-qc]} \BibitemShut
  {NoStop}%
\bibitem [{\citenamefont {Wald}(1993)}]{Wald:1993nt}%
  \BibitemOpen
  \bibfield  {author} {\bibinfo {author} {\bibfnamefont {R.~M.}\ \bibnamefont
  {Wald}},\ }\bibfield  {title} {\bibinfo {title} {{Black hole entropy is the
  Noether charge}},\ }\href {https://doi.org/10.1103/PhysRevD.48.R3427}
  {\bibfield  {journal} {\bibinfo  {journal} {Phys. Rev. D}\ }\textbf {\bibinfo
  {volume} {48}},\ \bibinfo {pages} {R3427} (\bibinfo {year} {1993})},\ \Eprint
  {https://arxiv.org/abs/gr-qc/9307038} {arXiv:gr-qc/9307038} \BibitemShut
  {NoStop}%
\bibitem [{\citenamefont {Iyer}\ and\ \citenamefont
  {Wald}(1994)}]{Iyer:1994ys}%
  \BibitemOpen
  \bibfield  {author} {\bibinfo {author} {\bibfnamefont {V.}~\bibnamefont
  {Iyer}}\ and\ \bibinfo {author} {\bibfnamefont {R.~M.}\ \bibnamefont
  {Wald}},\ }\bibfield  {title} {\bibinfo {title} {{Some properties of Noether
  charge and a proposal for dynamical black hole entropy}},\ }\href
  {https://doi.org/10.1103/PhysRevD.50.846} {\bibfield  {journal} {\bibinfo
  {journal} {Phys. Rev. D}\ }\textbf {\bibinfo {volume} {50}},\ \bibinfo
  {pages} {846} (\bibinfo {year} {1994})},\ \Eprint
  {https://arxiv.org/abs/gr-qc/9403028} {arXiv:gr-qc/9403028} \BibitemShut
  {NoStop}%
\bibitem [{Note3()}]{Note3}%
  \BibitemOpen
  \bibinfo {note} {In order to accomplish this, the topological Gauss-Bonnet
  term must be included in the action. Being topological it does not affect the
  equations of motion, but it contributes non-trivially to the
  entropy.}\BibitemShut {Stop}%
\end{thebibliography}%

\end{document}